\documentclass[journal]{IEEEtran}
\usepackage{cite}
\ifCLASSINFOpdf
  \usepackage[pdftex]{graphicx}
  \DeclareGraphicsExtensions{.pdf,.jpeg,.png}
\else
\fi
\usepackage{graphicx}
\usepackage{calc}
\usepackage{amsmath}
\usepackage{subfigure}
\begin{document}
\title{Intra-Frame Error Concealment Scheme using 3D Reversible Data Hiding in Mobile Cloud Environment}
% was not built to handle multiple paragraphsYi
%

\author{Yanli~Chen,~%\IEEEmembership{Non-Member,~IEEE,}
        Hongxia~Wang,~%\IEEEmembership{Fellow,~OSA,}
        Hanzhou~Wu,~%\IEEEmembership{Fellow,~OSA,}
       Yi Chen,~
       Asad Malik%\IEEEmembership{Life~Fellow,~IEEE}% <-this % stops a space
\thanks{Yanli Chen is with the school of Information Science and Technplogy,
Southwest Jiaotong University,Chengdu 611756,Sichuan,China,and also with
the School of Engineering, Tibet University, Lhasa 850000, China (e-mail:
\text{yanli\_027@163.com}).}% <-this % stops a space
\thanks{Hongxia~Wang, Yi Chen and Asad Malik are with the school of Information Science and Technplogy,
Southwest Jiaotong University,Chengdu 611756,Sichuan,China.}% <-this % stops a space
\thanks{Hanzhou Wu is Institute of Automation, Chinese Academy of Sciences (CAS), Beijing 100190, China.}
\thanks{Manuscript received April 19, 2005; revised August 26, 2015.}}

\maketitle

\begin{abstract}
Data in mobile cloud environment are mainly transmitted via wireless noisy channels, which may result in transmission errors with a high probability due to its unreliable connectivity. For video transmission, unreliable connectivity may cause significant degradation of the content. Improving or keeping video quality over lossy channel is therefore a very important research topic. Error concealment with data hiding (ECDH) is an effective way to conceal the errors introduced by channels. It can reduce error propagation between neighbor blocks/frames comparing with the methods exploiting temporal/spatial correlations. The existing video ECDH methods often embed the motion vectors (MVs) into the specific locations. Nevertheless, specific embedding locations cannot resist against random errors. To compensate the unreliable connectivity in mobile cloud environment, in this paper, we present a video ECDH scheme using 3D reversible data hiding (RDH), in which each MV is repeated multiple times, and the repeated MVs are embedded into different macroblocks (MBs) randomly. Though the multiple embedding requires more embedding space, satisfactory trade-off between the introduced distortion and the reconstructed video quality can be achieved by tuning the repeating times of the MVs. For random embedding, the lost probability of the MVs decreases rapidly, resulting in better error concealment performance. Experimental results show that the PSNR values gain about 5dB at least comparing with the existing ECDH methods. Meanwhile, the proposed method improves the video quality significantly.
\end{abstract}


% \fi
\IEEEpeerreviewmaketitle

\section{Introduction}
\IEEEPARstart{W}{ith} the rapid growth of information and communication technology, multimedia is becoming the popular format in the Internet. However, the storage space, computer resource, and bandwidth limit the development of multimedia communication. Cloud computing moves services, computation, and data to location-transparent centralized facilities or providers to solve these problems \cite{R1}.

The cloud services are becoming more and more popular, especially in the computing and storage aspects. Now, many multimedia companies have cloud-based platforms or shift their storage and computing service to the third parties, e.g., Youtube, Dailymotion, Tencent, and iqiyi. And the services provided by cloud can be used without high cost and complex architecture. In this case, with the development of terminal equipment, mobile cloud becomes a popular environment. However, for the unreliable connectivity of mobile cloud environment, it creates new problems for users. Mobile cloud architecture may bring out transmission degradation, especially  transmission errors in mobile cloud \cite{R2}.

Currently, video quality is becoming more attractive, especially for the medical video,surveillance video. To guarantee better video services, mobile cloud provides efficient platform for computing, storage and transmission. Since there is massive redundancy in the video sequence, they always need to be compressed before storing and transmitting. The popular video compression standard such as MPEG-2/4, H.264/AVC, and H.265 demands either parallel or distributed processing platform \cite{R2}. In the case of cloud computing, many Video Service Providers (VSPs) rent out the distribution architecture from Cloud Service Providers (CSPs) \cite{R3}.

The highly compressed videos are sensitive to the transmission errors. In order to ensure the quality, the video transmission encounters many challenges \cite{R2}: 1) Since both the low and high speed networks are existing in the Internet, it may cause data buffering at different locations during data transmission, then out of order delivery and packet loss or drops may occur; 2) Video transmission is based on connectionless protocols, and it cannot provide lossless data transmission.

For the video compression standard, there are two basic modes, inter-frame and intra-frame. Usually, for a frame, macroblocks (MBs) are the processing units under intra mode, and a processing unit may be a frame itself or a MB under inter mode. That is to say the video quality is mainly based on frame/MB quality. To improve the video quality, error resilience and error concealment are used as popular technologies. The error resilience techniques are used to recover the errors at the encoder side, and the error concealment techniques are used to conceal errors at the decoder side especially for the problem of the packet lost.

There are two methods for error concealment at the decoder side, data hiding-based method and the spatially/temporally correlation-based (data interpolation) method \cite{R1,R4}. The data interpolation method, which works well \cite{R5,R6} in the case of the successfully received data, can be used to reconstruct the lost data. For data hiding-based method, the performance depends on that whether the marked data can reflect the frames characteristics. Generally, the mark data consists of motion vectors (MVs) or residual which described the motion features.

Data hiding is a popular technique used in the field of multimedia security such as copyright protection, content authentication and secret communication \cite{R7}. Because the data hiding has the advantage that the mark data can be carried in a seamless way, it has been employed to conceal video errors produced by the transmission. Adsumilli \emph{et al.}\cite{R8} proposed an error concealment technique in which the 2-level discrete wavelet transforms (DWT) approximation coefficients are converted to halftone image which is embedded into original frame. Yilmaz \emph{et al.} \cite{R9} presented an error concealment method using edge orientation information as marked data. Ref. [4] compressed the residuals of the neighbor frames by compressed sensing (CS) as marked data at the encoder side, and reconstructed the original residual by CS to conceal the channel errors at decoder side.

In general, the more information about frame is embedded, the better error concealment performance. However, embedding more information means to degrade the quality of marked video. For the highly correlation between neighbor frames, the MVs of every MB are hidden to conceal the corresponding error MBs \cite{R10,R11}. In Yao \emph{et al.} \cite{R10}, the MVs of the MBs belonging to the region of interest (ROI), which are shared in a frame group, were embedded into the background region within the same frame. In Chen \emph{et al.} \cite{R11}, the MVs of every MB are embedded into the quantized discrete cosine transform (QDCT) coefficients of neighboring MBs, and at the decoder side, the extracted MVs are used to find the matching MBs. However, it modified the DCT coefficients which may lead to video quality degrade.

To strike balance between the marked video quality and error concealment performance, reversible data hiding (RDH) \cite{R12,R13} can recover the cover losslessly. Chung \emph{et al.} \cite{R14} presented a video error concealment scheme using RDH. Each MV is embedded into QDCT coefficients with zero values of the corresponding MB by a circular embedding scheme. Xu \emph{et al. }\cite{R15} proposed a RDH-based intra-frame error concealment method, in which it contains MV data pre-processing and the selection of embedding region. Later, Xu \emph{et al.} \cite{R7} raised a two dimensional RDH-based intra-frame error concealment scheme, and it is mainly used to reduce the distortions caused by RDH.

However, their performance needs to be improved for two reasons. Firstly, for the random of the channel state, errors are occurred randomly. So the error concealment performance in \cite{R14,R15,R16} cannot benefit from changeless embedding location. Secondly, for smooth video sequences, embedding MVs of the MBs sized $16\times16$ is a valuable selection to gain better error concealment performance and reduce the amount of marked data. However, to enhance the robustness of data hiding, the marked data may be produced by repeating MVs multiple times, and even more MVs with smaller block size are needed to be embedded into the frames. In order to solve these problems, an intra-frame error concealment scheme using 3D RDH is proposed. Since the mobile cloud-based services are contributed on the distributed transmitted system \cite{R2}, and almost all the client ends are connected to the network with wireless, the video can be transmitted effectively in mobile cloud environment in the case of limited bandwidth.

In this paper, to combine with the popular compression standard, H.264 is selected as the implementation platform. The rest of this paper is organized as follows. In Section 2, the motivation and novelty of proposed scheme is described. The main idea of 3D RDH is illustrated in Section3. And then, the proposed 3D RDH based error concealment method is presented in Section 4. In Section 5, experimental results are introduced. At last, the paper is concluded in Section 6.

\section{ Motivation and Novelty}

For a video transmitted through mobile cloud environment, the received video quality is mainly decided by the channel state in mobile cloud environment. Since the unreliable connectivity is existing in mobile cloud, for the video signal, error concealment methods are needed to improve the visual video quality. Meanwhile, there are much redundancy between the neighbor frames in a video. To raise video communication efficiency, the video signal is compressed before transmitting, and only the residuals between two frames/MBs are transmitted. So, some of error concealment methods are based on these temporally/spatially correlation. However, if the reference frames/MBs are corrupted, the errors may be propagated in the current frame/MB. Error concealment scheme using data hiding (ECDH) can conceal the errors from the extracted data, and its performance usually depends on that whether the mark data can reflect the frame/MB feature well. Then, a video ECDH is proposed in this paper to conceal the errors caused by the unreliable connectivity.

Secondly, at the decoder side, since the data hiding is realized by embedding data into QDCT coefficients. In other words, there are difference between the marked QDCT coefficients and the original QDCT coefficients. Also, these difference will produce high distortions in the decoded frames, and higher quantization steps make higher distortions. In order to avoid these distortions, a reversible data hiding method that the carrier can be recovered losslessly after data extraction is used in the proposed ECDH scheme.

Thirdly, for a MB, there is only one MV( include horizontal and vertical components, it is the same in the rest of this paper) that is hidden in the frame. If the MV is embedded into a corrupted MB, the MB cannot be concealed. To raise the probability of success concealment, every MV is repeated multiple times before being embedded, and these MV, belonging to a specified MB, are embedded into MBs in a frame separately and randomly. Then, if a MB is corrupted, only one MV is enough to reconstruct the MB. Also, a MB and its MV may be lost at the same time with lowest probability. However, MVs repeating makes the amount of marked data increasing. A 3D RDH is proposed to satisfy the demands of the embedding capacity-distortion performance, and then the video quality is improved. This also the novelty of this paper.

\section{ Proposed 3D RDH Scheme}
The methods in \cite{R14,R15} focus on using one-dimensional (1D) coefficient histogram for RDH. The 1D RDH is usually defined as \cite{R7}
\begin{eqnarray}\label{e1}
h(r)=\natural \{f_{i,j}(k)|f_{i,j}(k)=r\}.
\end{eqnarray}
where $\natural$ denotes the cardinal number of a set. Specifically, by considering every two adjacent residual coefficients together, 2D RDH can be defined as \cite{R7}
\begin{eqnarray}\label{e2}
h(r_{1},r_{2})=\natural \{f_{i,j}(2m),f_{i,j} (2m+1)|f_{i,j}(2m)=r_{1},f_{i,j} (2m+1)=r_{2}\}.
\end{eqnarray}

For 2D RDH, more than two bits may be embedded into a pair of residual coefficients \cite{R20}. Furthermore, bins '11' can be embedded by modifying only one coefficients \cite{R7,R20}. Obviously, the distortion is lower than that in 1D RDH before. However, they cannot satisfy the demand of the capacity-distortion performance. For 3D RDH method, every three adjacent residual coefficients are considered as a coefficient triple, it can be defined as
\begin{eqnarray}\label{e3}
h(r_{1},r_{2},r_{3} )=\natural\{(f_{i,j}(3m),f_{i,j}(3m+1),f_{i,j}(2m+3))|f_{i,j}(3m)=r_{1},\\\nonumber
f_{i,j}(3m+1)=r_{2},f_{i,j}(3m+2)=r_{3}\}~~~~~~~~~~~~~~~~~~~~~~~
\end{eqnarray}

For a coefficient triple, the transform space is more than that in 2D RDH. In 2D RDH, the two adjacent residual coefficients can be modified to four different pairs. However, in 3D RDH, a coefficient triple can be modified to eight different possible triples, and it can provide more capacity.

\begin{figure}
  \centering
  \includegraphics[width=9cm]{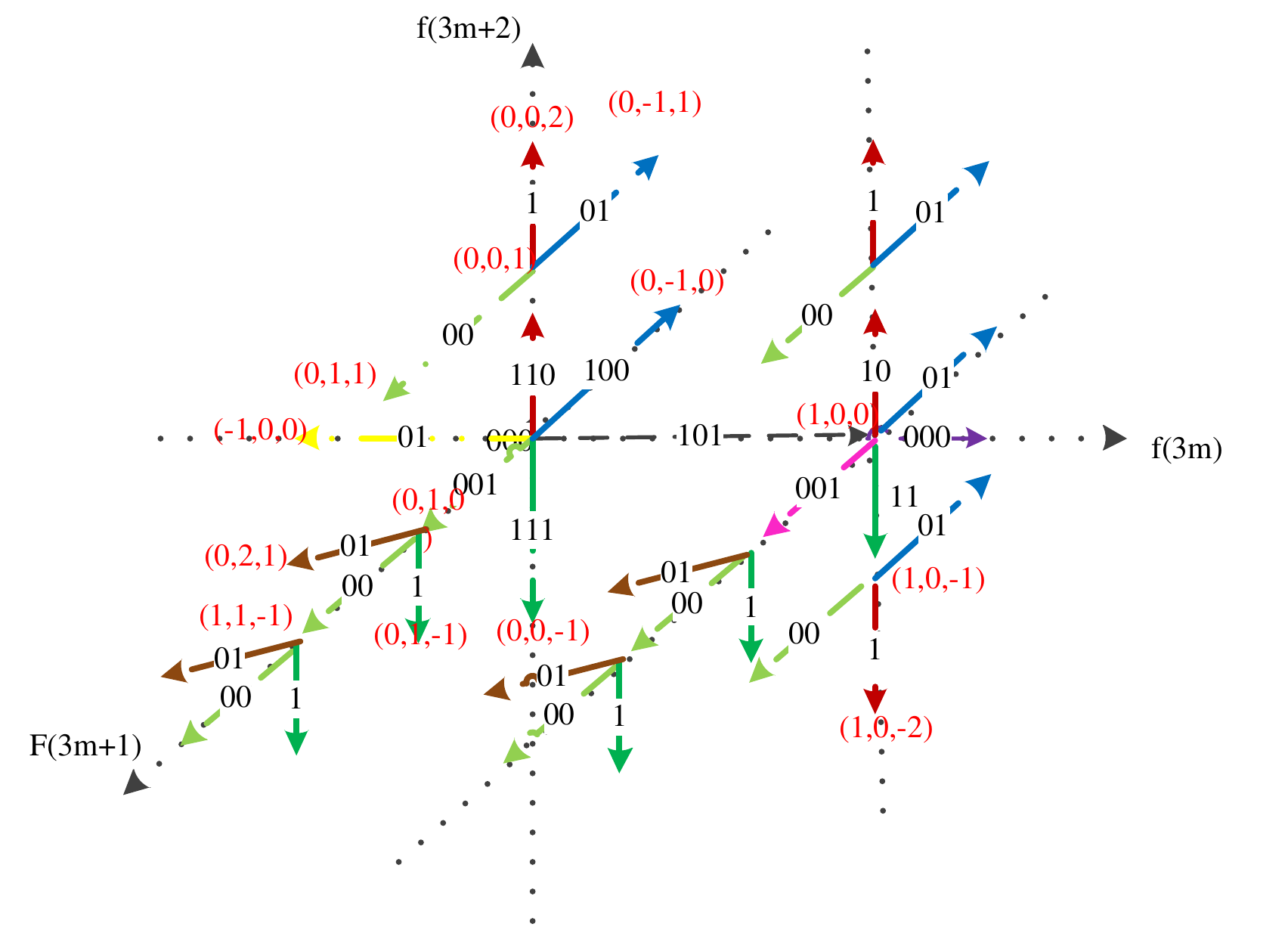}\\
  \caption{Illustration of the proposed 3D reversible data hiding}\label{3DRDH}
\end{figure}

Generally, the amount of the mark data is constant. However, most of the MVs have zero or near-zero values \cite{R7}. For 1D RDH \cite{R14,R15}, residual coefficients do not need to be expanded \cite{R7} in this occasion, but the histogram shifting operation is still to be carried out for the non-zero coefficients. For 2D RDH, beside the coordinate (0, 0), half coefficients in a coefficients pair are needed to be modified to embed one/two bits. In \cite{R7,R20}, at most two bits are embedded into a pair, and only one coefficient expands 1. So, both the 1D RDH and 2D RDH cannot meet the demands of the capacity-distortion performance well. In order to solve this problem, a 3D RDH with high capacity and low distortion is proposed and it is shown in Fig.\ref{3DRDH}.

Recently, a 3D RDH scheme is proposed in \cite{R21} to decrease the distortion. In \cite{R21} at most two bits are embedded into a coefficient triple, and it is not very suitable for our application scenarios. Our intention is to embed more bits and shift histogram in a less distorted direction as much as possible.
\section{Proposed error concealment scheme}
For the intra-frame error concealment, it mainly focus on the corrupted block. And for the inter-frame error concealment, it focus on the whole frame errors, such as delete or lost a frame, insert a frame, mis-order frames, and so on. In this scheme, we pay our attentions on utilizing 3D RDH and MV characteristics, which can reflect the most matching blocks, to conceal the corrupted blocks in a frame. The proposed 3D RDH in Section 3 is used to increase the embedding capacity and decrease the embedding distortions. The MVs are used as the secret data that are embedded in a frame, and also help the receiver to find the most matching blocks to conceal the corrupted blocks. If the sender want to communicate with the receiver, he/she will transport a secret key to the receiver, which is used to make the same scrambled MVs for the sender and receiver.
\subsection{Mark Data generation}

(1) Mark data

To provide high quality video, a high capacity RDH-based error concealment method is needed. All the schemes in \cite{R7,R14,R15} selected the MVs of $16\times16$ MBs as the marked data, and they can satisfy the demand of lower activity video sequence. To improve the robustness, the marked data is repeated multiple times, and more information may be embedded into the video sequence.

For a MB with size $16\times16$, if the search range is ㊣15, the total number of information bits to be embedded into a MB is \cite{R7}:
\begin{eqnarray}\label{e4}
L_{16}=2℅(\lceil log_{2} (2℅15+1)\rceil)=10
\end{eqnarray}
where $\lceil\rceil$ denotes the ceiling function. In this paper, to enhance the robustness of the marked data, at the encoder side, all the MVs in a frame are repeated multiple times as marked data before being embedded. Assuming every MV is repeated $\alpha$ times, then the total number of information bits to be embedded into one MB is:
\begin{eqnarray}\label{e5}
L=\alpha\cdot L_{16}
\end{eqnarray}
                           	
 Significantly, with the $\alpha$ increasing, a large capacity RDH scheme is needed, and the embedding capacity is controllable by adjusting $\alpha$.

(2) Embedding locations

In  \cite{R7}, \cite{R14} and \cite{R15}, the MV of every MB is embedded into the neighbor MB. However, in actuality, in mobile cloud environment, the errors may be spread multiple neighbor blocks in a higher probability, the stationary embedding locations cannot resist the random channel errors.

In \cite{R7,R10,R11,R14,R15,R16}, every $4\times4$ neighbor MBs are considered as a MB group, MVs are embedded into the neighbor MBs circularly. There are horizontal, vertical and diagonal patterns, and only vertical pattern can recover two MBs \cite{R15}. However, for mobile cloud environment, the errors are occurred randomly, and the neighbor MBs may be corrupted in the same time. To keep consistent with the random channel, all the repeated MVs in a frame are scrambled randomly, then every $\alpha$  MVs are considered as a group and embedded into $\alpha$ MBs sequentially. In other words, the $\alpha$  MVs belonging to the same MB may be distributed in $\alpha$ different MBs randomly. And for a lost MB, it is enough to conceal the error that only one of the 汐 repeated MVs is embedded into the non-corrupted MB. Meanwhile, The host MBs and their all repeated MVs are lost at the same time with a much lower probability. Also, the repeating frequency can be adjusted according to the demands of the video application.

In order to illustrate the performance of random embedding MVs further, assuming there are N blocks in a frame, and the PLR is $p$. For a specified MB denoted as B, it is lost with probability $p$. Since the repeated MVs are scrambled randomly, the repeated MVs belonging to the specified MB may be embedded into $\beta(\beta=1,2,\cdots,\alpha)$ MBs. If we do not consider the influence from other MVs in the embedding process, the $\alpha$ MVs are embedded into $\beta$ MBs with probability $P_{\beta}=\dbinom{N}{\beta}/\sum_{j=1}^{\alpha}\dbinom{N}{\beta}$ . Furtherly, in the embedding process, the worst condition is that there is only one location to embed the $\alpha$ MVs, then $P_{\beta}=1$. If the $\beta$ host MBs are corrupted at the same time, MB B cannot be concealed correctly. This case occurs with probability $P_{nc}=\sum_{\beta=1}^{\alpha}P_{\beta}\cdot p^{\beta}$, and it is also the probability that a MB cannot be concealed correctly. Conversely, for a lost MB, even if only one of its $\alpha$ MVs is embedded in the normal MB, MB B can be concealed correctly. So a lost MB can be concealed with the probability $P_{c}=1-P_{nc}$. Obviously, under the normal channel, $p<0.5$, the relationship between $P_{c}$ and $P_{nc}$ is described as
\begin{eqnarray}\label{e6}
P_{c}=1-P_{nc}>P_{nc}
\end{eqnarray}
   	
\begin{figure}
  \centering
  \includegraphics[width=8cm]{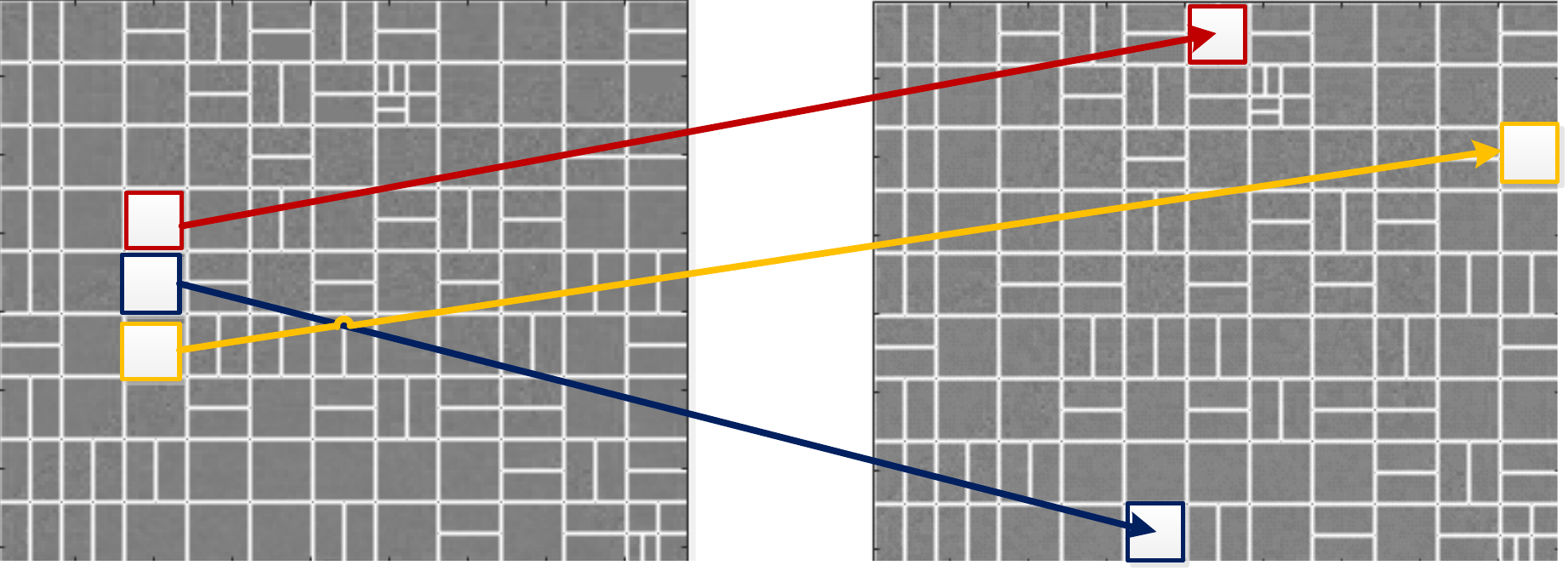}\\
  \caption{The mapping of host MB and scrambled MVs with $\alpha=1$}\label{MVmapping1}
\end{figure}

\begin{figure}
  \centering
  \includegraphics[width=8cm]{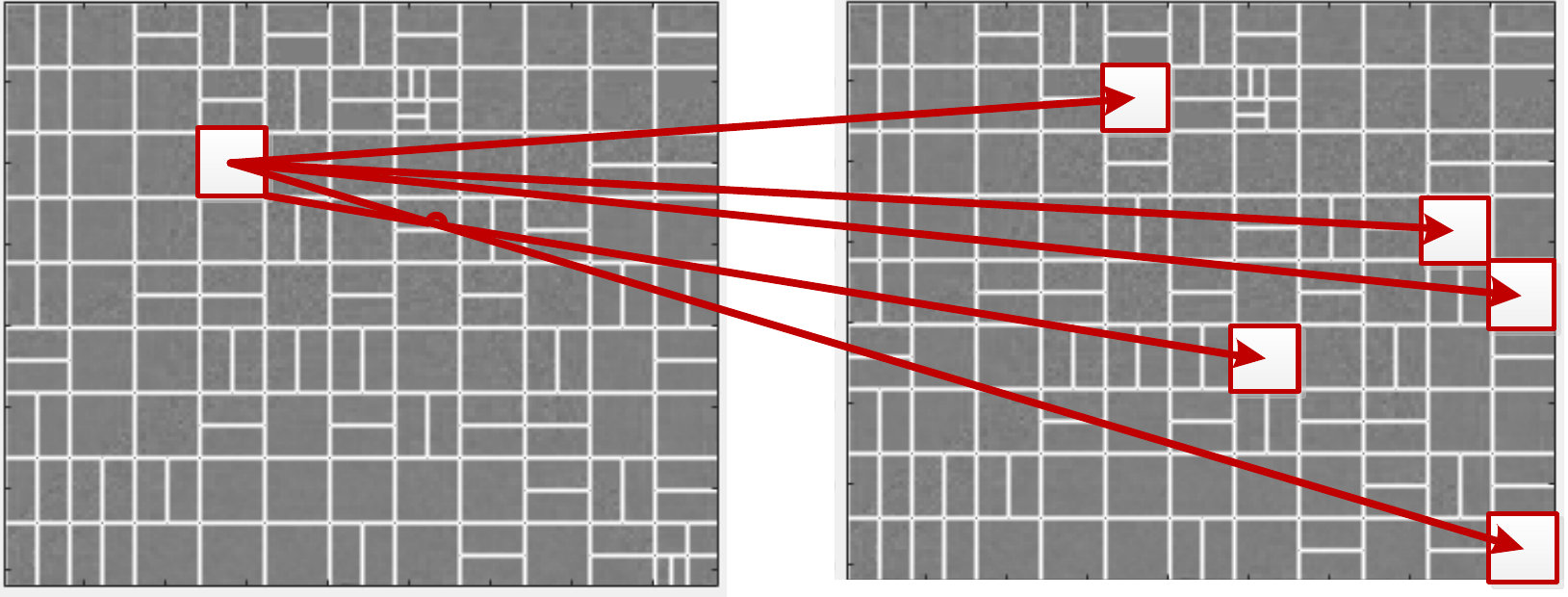}\\
  \caption{The mapping of host MB and scrambled MVs with $\alpha=5$}\label{MVmapping5}
\end{figure}

For the random embedding location, the relationship between host MBs and the embedding locations of their MVs is changeable. Fig.\ref{MVmapping1} and Fig.\ref{MVmapping5} shows the mapping of host MBs and the embedding locations of their MVs in the residual coefficients frame in video ※hall§  with $\alpha=1$ and $\alpha=5$, and it shows the neighbor MBs in the left(the original host MBs) are scrambled dispersedly in the right residual coefficients frame (scrambled embedding locations).

\begin{figure}
  \centering
  \subfigure[PLR=0.1]{
    %\label{fig:subfig:a} %% label for first subfigure
    \includegraphics[width=5cm]{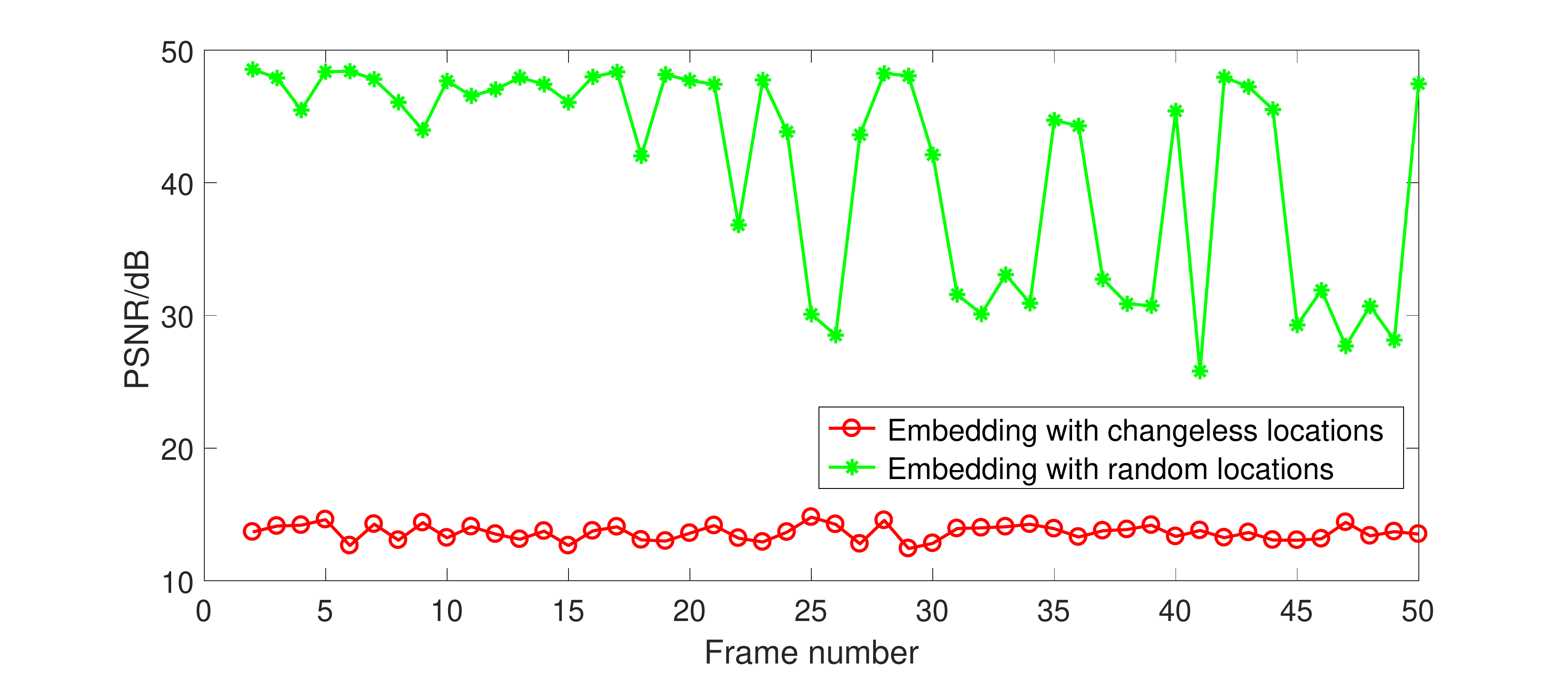}}
  %\hspace{1in}
  \subfigure[PLR=0.3]{
   % \label{fig:subfig:b} %% label for second subfigure
    \includegraphics[width=5cm]{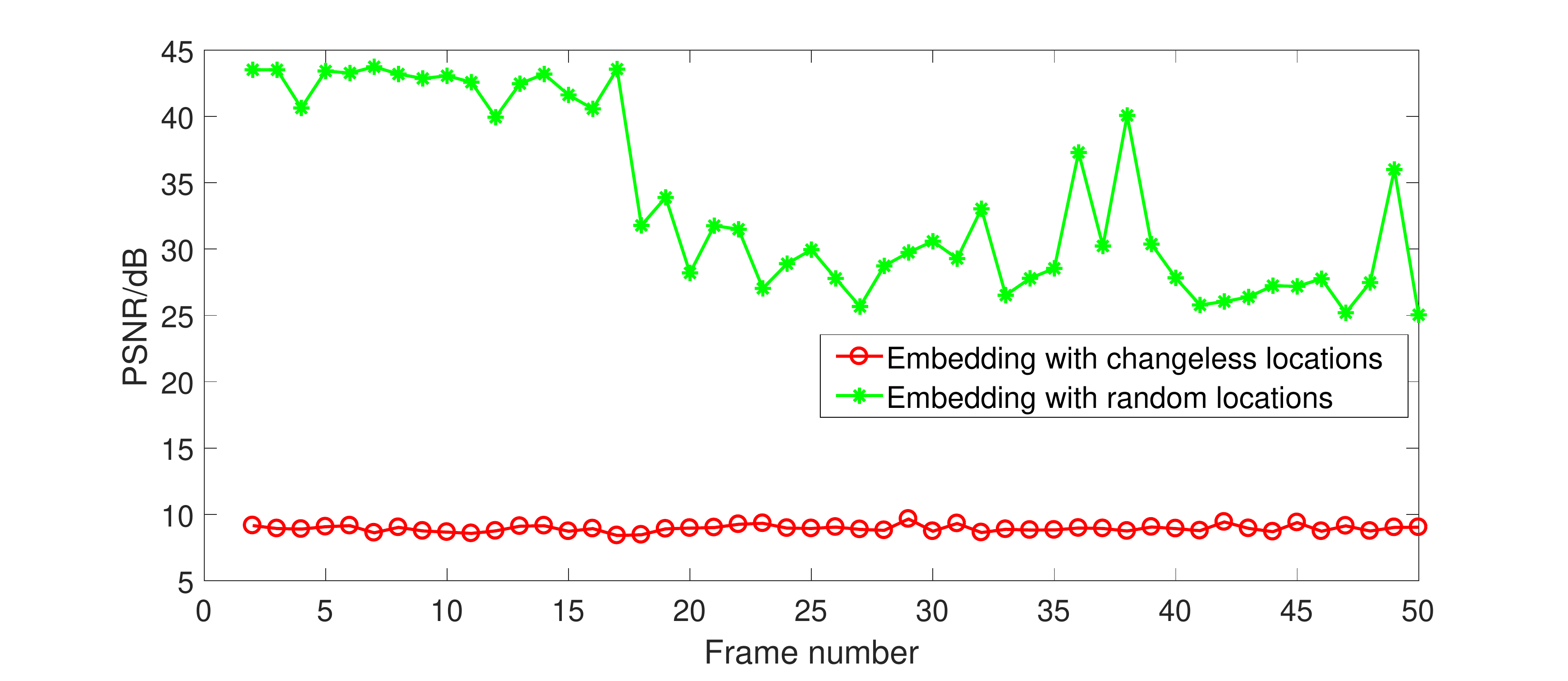}}
  \caption{PSNR of concealed video sequences foreman with different PLR: (a) PLR=0.1 ;( b) PLR=0.3}
  \label{random}
\end{figure}

In order to show the influence on concealment performance significantly, in this paper, all the corrupted MBs that cannot be concealed correctly are set to be zeros and all the concealment experimental results in this paper are implemented under this case. In the other words, if the MBs and their MVs are lost at the same time, the decoded MBs are substituted by black blocks. To reveal the impact from random embedding on the concealment performance, Fig.\ref{random} shows the comparison of concealed frames PSNR between the random embedding and changeless embedding with PLR 0.1, 0.3 and $\alpha=1$, it reveals that random embedding can achieve a better performance.

At the decoder side, for a MB, there would be multiple same MVs for a specified MB, and they may be extracted from different MBs. So, if a MB is corrupted, it can be concealed in a high probability.
\begin{figure}
  \centering
  \includegraphics[width=8cm]{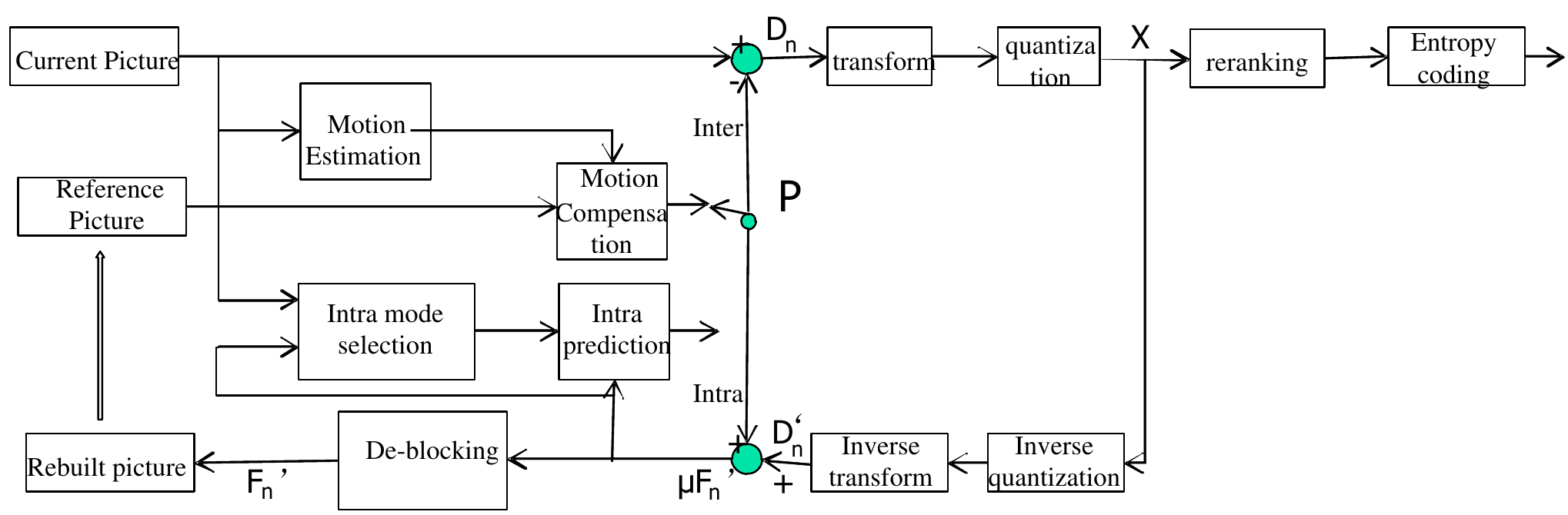}\\
  \caption{The H.264 Encoding framework}\label{H264}
\end{figure}
\subsection{Data Hiding}

Fig.\ref{H264} shows the H.264 encoding procedure \cite{R18,R19}. Considering the video compressing process, the marked data is embedded into the quantized DCT coefficients, and the mid-frequency and high-frequency coefficients are used to carry the MVs. For different repeating frequency requires different embedding capacity, the AC component is reserved for embedding. Since the smallest encoding unit is $4\times4$ block, all the AC components of QDCT coefficients are considered as the cover. After zigzag scanned, every three adjacent coefficients are grouped to be a triple. After that, the mid-frequency and high-frequency coefficients are divided into triples in order. The modifications on these triples can be actualized based on the proposed 3D RDH in Section 3. Every embedding operation is actualized by modifying one or two elements in a triple with $\pm1$. The original coefficients triple $(f_{i,j}(3m),f_{i,j}(3m+1), f_{i,j} (3m+2))$, the marked coefficients triple $(f_{i,j}^{'}(3m),f_{i,j}^{'}(3m),f_{i,j}^{'}(3m))$ and the recovered coefficient triple $(f_{i,j}^{"}(3m),f_{i,j}^{"}(3m),f_{i,j}^{"}(3m))$ are denoted as $F,F^{'},F^{"}$.

The embedding procedure is as follows.
	
(1) If $F=(0,0,0)$, the marked coefficient triple $F^{'}$ is determined as follows
\begin{eqnarray}\label{e7}
F^{'}=\left\{
                        \begin{array}{rc}
                          (0,0,0)~~   ~~~&if~~ b_{1}b_{2}b_{3}=000\\
                          (0,1,0)~~  ~~~&if ~~b_{1}b_{2}b_{3}=001\\
                          (-1,0,0)  ~~~&if~~ b_{1}b_{2}  = 01~~~~\\
                          (0,-1,0)  ~~~&if~~ b_{1}b_{2}b_{3}= 100\\
                          (1,0,0)~~   ~~~&if~~ b_{1}b_{2}b_{3}= 101\\
                          (0,0,1)~~    ~~~&if~~ b_{1}b_{2}b_{3}= 110\\
                          (0,0,-1)  ~~~ &if~~ b_{1}b_{2}b_{3}= 111
                        \end{array}
                      \right..
\end{eqnarray}

where $b_{i},i=1,2,\cdots$ denotes the to-be-embedded bits. In this case, three bits are embedded by modifying one element in a triple at most with the probability 0.75.

(2) If $F=(x,0,z),z\neq0$, the marked coefficient triple $F^{'}$ is determined as follows
\begin{eqnarray}\label{e8}
F^{'}=\left\{
                        \begin{array}{rc}
                          (x,1,y)  ~ ~~~~~~~~~~~~& if~~ b_{1}b_{2}=00\\
                          (x,-1,y)  ~~~~~~ ~~~~~ &if ~~b_{1}b_{2}=01\\
                          (x,0,y+sign(y))  &if ~~b_{1}=1~~~~
                        \end{array}
                      \right..
\end{eqnarray}
where $sign(y)$ denotes the symbol of $y$ .In this case, at least one bit is embedded with modifying one of three coefficients. Comparing with the 3D RDH \cite{R21}, the capacity-distortion performance is improved.
	
(3) If $F=(x,0,0),x\neq0$, the marked coefficient triple $F^{'}$  is determined as follows
\begin{eqnarray}\label{e9}
F^{'}=\left\{
                        \begin{array}{rc}
                          (x+1,0,0)  &if~ b_{1}b_{2}b_{3}=000\\
                          (x,1,0)~~~~~    & if~ b_{1}b_{2}b_{3}=001\\
                          (x,-1,0)~~~   &if~ b_{1}b_{2}=01~~~~
                        \end{array}
                      \right..
\end{eqnarray}

In this case, more than two bits are embedded by expanding 1. However, in the 3D RDH [21], only one bit can be embedded with the same distortion.

(4) If $F=(x,y,0),x\neq0,y\neq0$, the marked coefficient triple $F^{'}$ is determined as follows
\begin{eqnarray}\label{e10}
F^{'}=\left\{
                        \begin{array}{rc}
                          (x,y+sign(y),0)   &if~~ b_{1}b_{2}=00\\
                          (x,y+sign(y),1)  &if~~ b_{1}b_{2}=01\\
                          (x,y,-1) ~~~~~~~~~~~ &if~~ b_{1}=1~~~~
                        \end{array}
                      \right..
\end{eqnarray}

In this case, one or two bit is embedded with modifying one coefficient. For the large quantization step, most of the AC coefficients are zeros. This case is occurred with low probability infrequently.

(5) If $f_{i,j}(2m+1)>0$ and $f_{i,j}(2m+2)\neq0$, the coefficient triple $F^{'}$  is shifted as follows.
\begin{eqnarray}\label{e11}
 F^{'}= (f_{i,j}(3m),f_{i,j}(3m+1)+1,f_{i,j}(3m+2)\\
 +sign(f_{i,j}(2m+2)))~~~~~~~~~~~~~~~~~~~~~\nonumber
 \end{eqnarray}

(6) If $f_{i,j}(2m+1)<0$ and $f_{i,j}(2m+2)\neq0$, the coefficient triple $F^{'}$ is shifted as follows.
\begin{eqnarray}\label{e12}
 F^{'}= (f_{i,j}(3m),f_{i,j}(3m+1)-1,f_{i,j}(3m+2)\\
 +sign(f_{i,j}(2m+2)))~~~~~~~~~~~~~~~~~~~~~~\nonumber
 \end{eqnarray}

In the case 5) and 6), all the shifting are realized by modifying two coefficients in a triple.
\subsection{Data Extraction and Original Video Recovery}

The original video can be recovered by extracting the hidden MVs according to the proposed 3D RDH in this paper, and error MBs can be concealed by replacing the matching MBs in reference frame. The data extraction process is actualized before de-quantization, so the host MBs can be decoded without any extra distortion when there are no transmitting errors. Once the transmitting errors occur, the error concealment process is realized after the data extraction.

Assuming there are no errors in the transmission, data extraction procedure can be described as follows

(1) If the marked coefficients triple $F^{'}\in\{(0,0,0),(1,0,0),(0,1,0),(0,0,1),\\
(-1,0,0),(0,-1,0),(0,0,-1)\},$ the hidden bits can be extracted as
\begin{eqnarray}\label{e13}
 bs=\left\{
                        \begin{array}{rc}
                          000   &if F^{'}=(0,0,0)~\\
                          001   &if F^{'}=(0,1,0)~\\
                          01~  &if F^{'}=(-1,0,0)\\
                          100   &if F^{'}=(0,-1,0)\\
                          110   &if F^{'}=(0,0,1)~\\
                          111   &if F^{'}=(0,0,-1)
                        \end{array}
                      \right..
 \end{eqnarray}

After the bits are extracted, the recovered coefficients triple $F^{"}$ recovers to (0,0,0).

(2) If the marked coefficients triple $F^{'}\in\{\{(x,0,z)||z|>1\},\{(x,1,z)||z|>0\},\{(x,-1,z)||z|>0\}\}$, the hidden bits are extracted as follows.
\begin{eqnarray}\label{e14}
 bs=\left\{
                        \begin{array}{rc}
                          00, &if F^{'}\in\{(x,1,z)||z|>0\}~~\\
                          01,&if  F^{'}\in\{(x,-1,z)||z|>0\}\\
                          1,~&if  F^{'}\in\{(x,0,z)||z|>1\}~
                        \end{array}
                      \right..
 \end{eqnarray}
After the bits are extracted, the host coefficient triples are recovered as
\begin{eqnarray}\label{e15}
 F^{"}=\left\{
                        \begin{array}{rc}
                         (x,0,z-sign(z)),&if F^{'}﹋\{(x,0,z)||z|>1\}\\
                          (x,0,z)~~~~~~~~~~~~,   & ~~~~~~others~~~~~~~~
                        \end{array}
                      \right..
 \end{eqnarray}

(3) If the marked coefficients triple$F^{'}  \in\{\{(x,0,0)||x|>1\},\{(x,0,1)|x\neq0\},\{(x,0,-1)|x\neq0\},\{(x,1,0)|x\neq0\},\{(x,-1,0)|x\neq0\}\}$, the hidden bits are extracted as
\begin{eqnarray}\label{e16}
 bs=\left\{
                        \begin{array}{rc}
                         000,&if  F^{'}\in\{(x,0,0)||x|>1\}\\
                         001,&if  F^{'}\in\{(x,1,0)|x\neq0\}~~\\
                         01~,&if  F^{'}\in\{(x,-1,0)|x\neq0\}\\
                         10~,&if  F^{'}\in\{(x,0,1)|x\neq0\}~~\\
                         11~,&if  F^{'}\in\{(x,0,-1)|x\neq0\}
                        \end{array}
                      \right..
 \end{eqnarray}

After the bits are extracted, the host coefficient triples are recovered as
\begin{eqnarray}\label{e17}
 F^{"}=\left\{
                        \begin{array}{rc}
                         (x-sign(x),0,0),&if F^{'}\in\{(x,0,0)||x|>1\}\\
                          (x,0,0),~~~~~~~~~~~~ & ~~~~~~~~~~~others~~~~~~~~~
                        \end{array}
                      \right..
 \end{eqnarray}

(4) If the marked coefficients triple $F^{'}\in\{\{(x,y,0)||y|>1\},\{(x,y,1)||y|>1\},\{(x,y,-1)||y|>0\}\}$, the hidden bits are extracted as
\begin{eqnarray}\label{e18}
 bs=\left\{
                        \begin{array}{rc}
                         00,&if F^{'}\in\{(x,y,0)||y|>1\}~\\
                         01,&if  F^{'}\in\{(x,y,1)||y|>1\}~\\
                         1~,&if  F^{'}\in\{(x,y,-1)||y|>0\}
                        \end{array}
                      \right..
 \end{eqnarray}

After the bits are extracted, the host coefficient triples are recovered as
\begin{eqnarray}\label{e19}
 F^{"}=\left\{
                        \begin{array}{rc}
                         (x,y,0),~~~~~~~~~~~~&if F^{'}\in\{(x,y,-1)||y|>0\}\\
                          (x,y-sign(y),0), & ~~~~~~~~~~~others~~~~~~~~~~~
                        \end{array}
                      \right..
 \end{eqnarray}

(5) If the marked coefficient triples belong to shifting triples, they are recovered by inverse-processing of shifting.

Unfortunately, if there are transmission errors, after all the bits are extracted, the MVs are obtained by inverse-scrambling the extracted MVs. The error MBs are concealed by replacing those using matching blocks.
\section{Experimental Result }

The proposed scheme is applied in the H.264, and the front 30 frames of six standard sequences in QCIF format ($176\times144$) are used to test the performance of the proposed scheme. Every 10 frames are considered as a GOP. In order to reveal the advantage of the proposed scheme, recent works in \cite{R7} and \cite{R15} are used to be compared with the proposed scheme.
\subsection{Embedding Distortion}

In this scheme, 3D RDH is used to embed the MVs into the quantized DCT residual coefficients. The goal of proposed RDH scheme is to improve embedding capacity and introduce low embedding distortion. However, since MVs are embedded into the host residual coefficients, it would cause difference between the host coefficients and marked coefficients. For the RDH algorithm, the host coefficients can be recovered completely from the embedding distortions if the video sequences are transmitted without any channel errors.

(1) ECDR

Absolute embedding capacity and distortion ratio (ECDR) can be defined simply as
\begin{eqnarray}\label{e20}
 ECDR=EC/D
 \end{eqnarray}
where $EC$ denotes the  absolute embedding capacity, $D$ denotes the absolute distortion, and both of them are calculated with the full embedding situation. In order to analyze the proposed 3D RDH performance better, the residual coefficients teiples are divided into multiple sets.  $C_{1},C_{2},C_{3}$  and $C_{4}$ denote the original coefficient triples sets of the 1st, 2nd, 3rd ,4rd case in Section 4, $C_{4}$ is defined in Eq.(\ref{e25}), $C_{5}$ denotes the original coefficient triples set of 5th and 6th case. The ECDR of the proposed 3D RDH is evaluated by comparing recent 3D RDH in \cite{R21}. Meanwhile, the embedding capacity of the proposed 3D RDH, and that in \cite{R21} are denoted as $EC_{pro},EC_{rec}$, and the distortions in these schemes are denoted as $D_{pro},D_{rec}$.
\begin{eqnarray}\label{e21}
 EC_{pro}=\frac{11}{4}\sharp C_{1}+\frac{3}{2}\sharp C_{2}+\frac{9}{4}\sharp C_{3}+\frac{3}{2}\sharp C_{4}
 \end{eqnarray}
 \begin{eqnarray}\label{e22}
 EC_{rec}=\frac{9}{4}\sharp C_{1}+\sharp C_{2}+2\sharp C_{3}+\sharp C_{4}
 \end{eqnarray}
 \begin{eqnarray}\label{e23}
 D_{pro}=\frac{7}{8}\sharp C_{1}+\sharp C_{2}+2\sharp C_{3}+\frac{5}{4}\sharp C_{4}+2\sharp C_{5}
 \end{eqnarray}
  \begin{eqnarray}\label{e24}
 D_{rec}=\frac{3}{4}\sharp C_{1}+\frac{3}{2}\sharp C_{2}+\frac{5}{4}\sharp C_{3} +\frac{3}{2}\sharp\{(x,y,0)||x|>2,|y|>0\}~~~~~~~~~~~~~~~~~~~~~~~~~\\\nonumber +2\sharp\{(x,y,0)||x|=1,|y|>0\} +\sharp\{(0,y,0)||y|>0\}+2\sharp C_{5}+\frac{3}{2}\sharp \{(0,0,1)\}\nonumber
 \end{eqnarray}
where $\sharp$ denotes the elements number in set. The embedding capacity and distortion difference are calculated as
 \begin{eqnarray}\label{e25}
 EC_{pro}-EC_{rec}=\frac{1}{2}\sharp C_{1}+\frac{1}{2}\sharp C_{2}+\frac{1}{4}\sharp C_{3}+\frac{1}{2}\sharp C_{4}
 \end{eqnarray} 	
   \begin{eqnarray}\label{e26}
 D_{pro}-D_{rec}=\frac{1}{8}\sharp C_{1}-\frac{1}{2}\sharp C_{2}+\frac{3}{4}\sharp C_{3}-\frac{1}{4}\sharp\{(x,y,0)||x|>2,|y|>0\}~~~~~~\\\nonumber
~~~~~~~~~~~~~~~~~~~~+\frac{1}{4}\sharp\{(0,y,0)||y|>0\}-\frac{1}{4}\sharp \{(0,0,1)\}~~~~~~~~~~~~~~~~~~~~~~~~~
 \end{eqnarray}

 Eq. (\ref{e25}) indicates that the embedding capacity in proposed scheme is much more than that in the scheme \cite{R21} and Eq. (\ref{e26}) indicates that the distortion difference in the two schemes is decided by the elements number in $C_{1},C_{2},C_{3}$ and $C_{4}$. However, for video compressed standard, the number of the  elements in these sets is mainly decided by quantization step. With the quantization step decreasing, coefficients with zero values decrease, and also the distortion difference may decrease. Generally, to guarantee the quality of the decoded sequence, the quantization step cannot be a number which make all the AC components are zeros. So the ECDR in the proposed scheme is more than that in the scheme \cite{R21} in great probability. Let $P(D_{pro}-D_{rec}>0)$ and $P(D_{pro}-D_{rec}<0)$ denote the probability of the case $D_{pro}-D_{rec}>0$ and $D_{pro}-D_{rec}<0$, then the relation of probability is described as
 \begin{eqnarray}\label{e27}
 P(D_{pro}-D_{rec}>0)<P(D_{pro}-D_{rec}<0)
 \end{eqnarray}

 Assuming $ECDR_{pro}$ and $ECDR_{rec}$ denote the ECDR in the proposed scheme and in \cite{R21}. According to Eq.(\ref{e25}) and Eq.(\ref{e26}), $ECDR_{pro}$ is greater than $ECDR_{rec}$ in a higher probability, and it can be described as follows.
 \begin{eqnarray}\label{e28}
 P(ECDR_{pro}=\frac{EC_{pro}}{EC_pro}>ECDR_{rec}=\frac{EC_{rec}}{EC_rec})>0.5
 \end{eqnarray}

\begin{figure}[ht]
\centering
\subfigure[coastguard]{%
\includegraphics[width=.45\linewidth - 0.25mm]{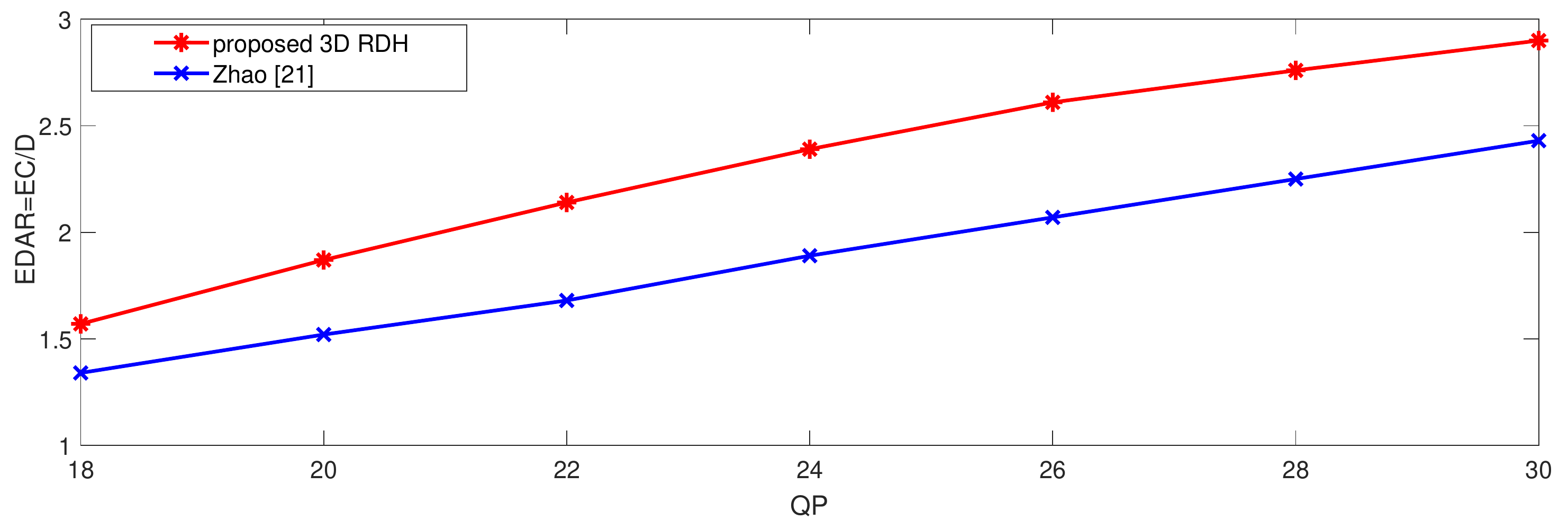}\hfill
\label{fig:subfigure1}}
\quad
\subfigure[foreman]{%
 \includegraphics[width=.45\linewidth - 0.25mm]{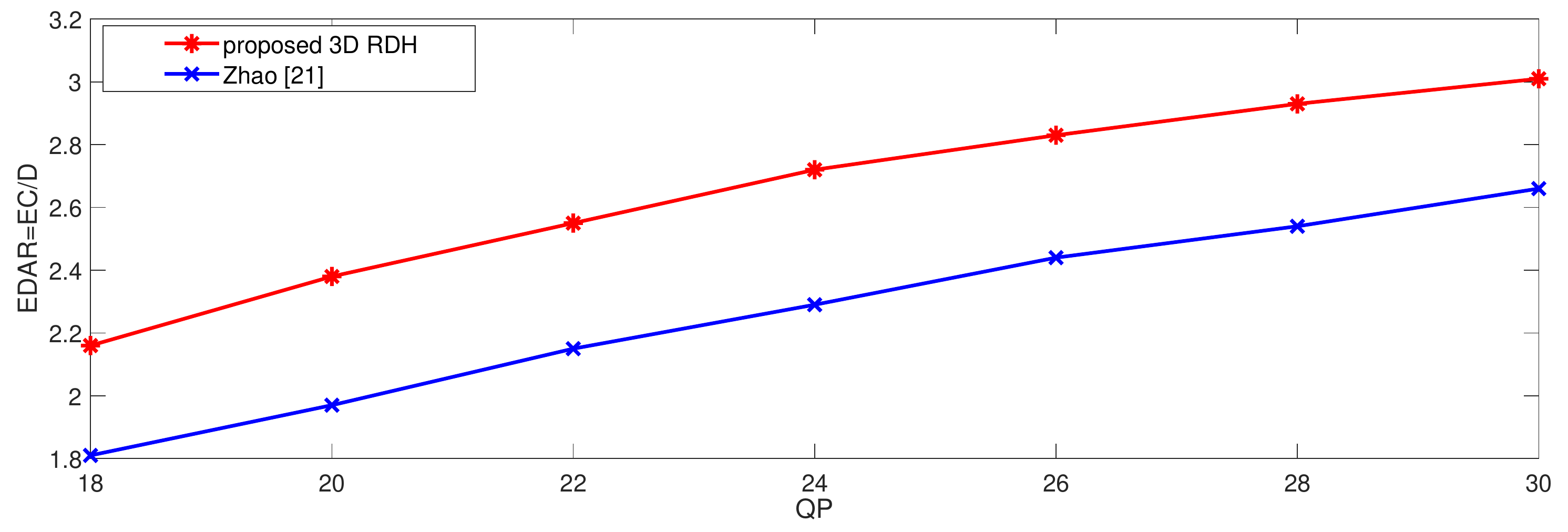}
\label{fig:foreman}}
\subfigure[hall]{%
\includegraphics[width=.45\linewidth - 0.25mm]{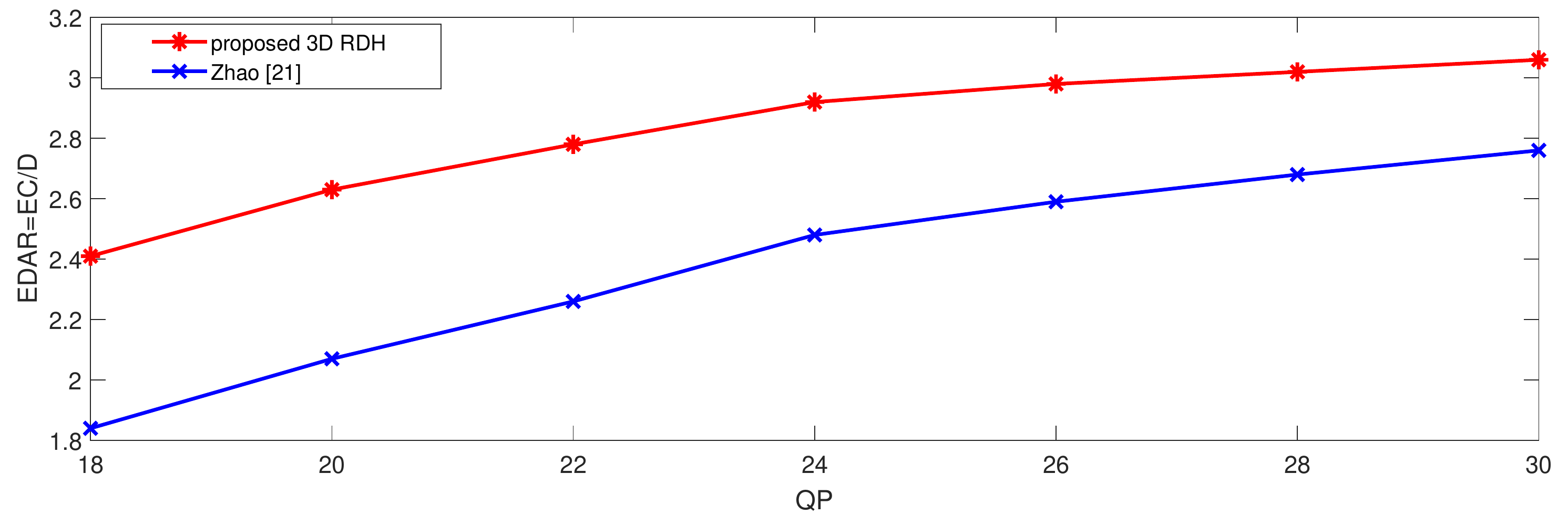}\hfill
\label{fig:hall}}
\quad
\subfigure[grandma]{%
  \includegraphics[width=.45\linewidth - 0.25mm]{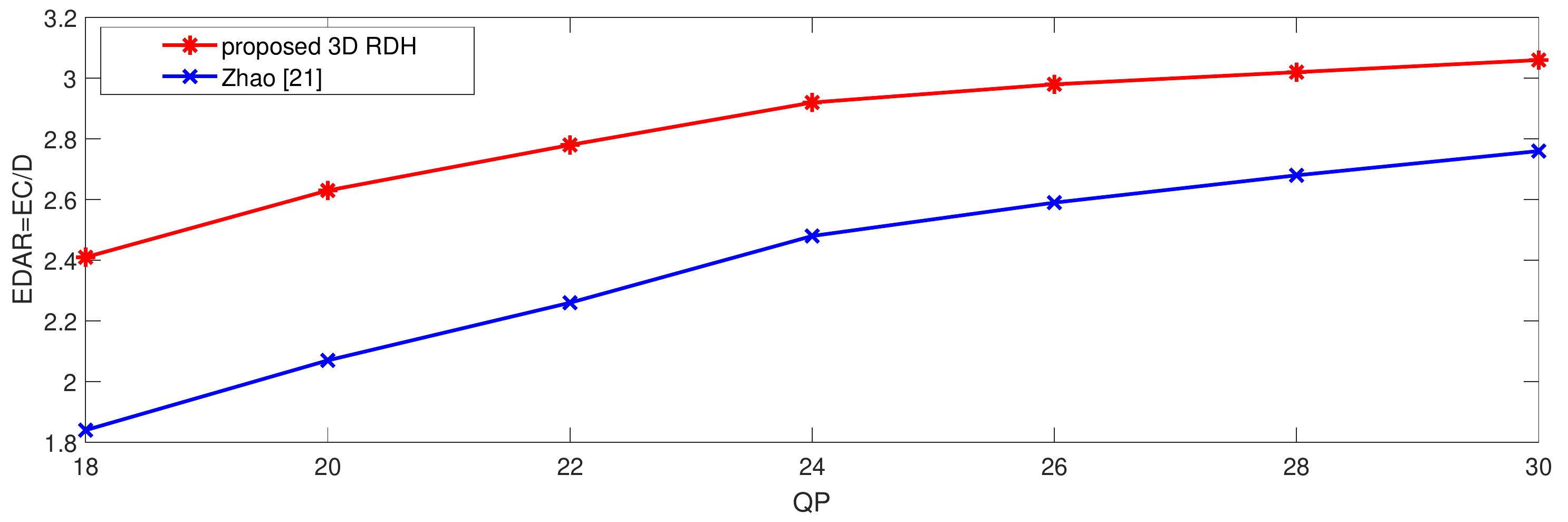}
\label{fig:grandma}}
\caption{The EDAR of proposed 3D RDH comparison with [21]}
\label{edar}
\end{figure}

 To illustrate the ECDR of the two 3D RDH scheme, emebedding capacity(EC) and the corresponding distortion of the 2nd frame in video sequence $foreman$, $hall$, $coastguard$ and $grandma$ are calculated in Table \ref{table1}. It shows that with the quantization step (QP) increasing, the embedding capacity is increased too, and the distortion is decreased for the decreased shifting operations with lower QP. It also shows the proposed 3DRDH can provide more embedding space.  Fig.\ref{edar} shows the ECDR comparison of the ECDR in proposed 3D RDH is nearly two times as that in [21].
 \begin{table*}\footnotesize
  \centering
  \caption{the embedding capacity and distortion comparison with \cite{R21}}\label{table1}
  \begin{tabular}{lccccccccr}
  \hline
   ~~~~~~~~QP~~~~~~~~  &18      &20     &22     &24     &26     &28     &30\\
  ~~~~~~~~~~~~~~~EC~~~proposed  &15802	&16981	&17939	&18732	&19444	&19994	&20484\\
  ~~~~~~~~~~~~~~~~~~~~~~~~[21]~~~~  &12792	&13807	&14633	&15295	&15905	&16358	&16762\\
  foreman~~~~~~D~~~~proposed  &7329	&7140	&7023	&6892	&6859	&6817	&6795\\
  ~~~~~~~~~~~~~~~~~~~~~~~~[21]~~~~  &7051	&7024	&6819	&6693	&6510	&6451	&6309\\
  \hline
  ~~~~~~~~~~~~~~~EC~~~proposed  &16694	&17893	&18771	&19822	&20225	&10577	&20875\\
  ~~~~~~~~~~~~~~~~~~~~~~~~[21]~~~~  &13560	&14595	&15336	&16204	&16537	&16828	&17072\\
  ~~hall~~~~~~~~~D~~~~proposed  &6920	&6816	&6758	&6787	&6797	&6816	&6817\\
  ~~~~~~~~~~~~~~~~~~~~~~~~[21]~~~~  &7372	&7048	&6795	&6528	&6383	&6280	&6186\\
   \hline
  ~~~~~~~~~~~~~~~EC~~~proposed  &12586	&14107	&15405	&16762	&17900	&18760	&19624\\
  coastguard~~~~~~~~~~~~~~[21]~~~~  &10034	&11350	&12459	&13625	&14591	&15320	&16041\\
  ~~~~~~~~~~~~~~~D~~~~proposed  &7996	&7539	&7190	&7002	&6857	&6807	&6758\\
  ~~~~~~~~~~~~~~~~~~~~~~~~[21]~~~~  &7480	&7472	&7411	&7195	&7043	&6798	&6589\\
  \hline
   ~~~~~~~~~~~~~~EC~~~proposed  &18781	&19342	&19730	&20398	&20679	&20925	&21254\\
  ~~grandma~~~~~~~~~~~~~~~~[21]~~~~  &15301	&15784	&16114	&16684	&16909	&17121	&17396\\
  ~~~~~~~~~~~~~~~D~~~~proposed  &6748	&6747	&6721	&6763	&6786	&6814	&6852\\
  ~~~~~~~~~~~~~~~~~~~~~~~~~[21]~~~~  &6871	&6738	&6557	&6361	&6276	&6179	&6100\\
  \hline
  \end{tabular}
  %\caption{Performance for Different Threshold}%\label{}
\end{table*}

(2) PSNRs of marked frames

\begin{figure}[ht]
\centering
\subfigure[grandma]{%
\includegraphics[width=.45\linewidth - 0.25mm]{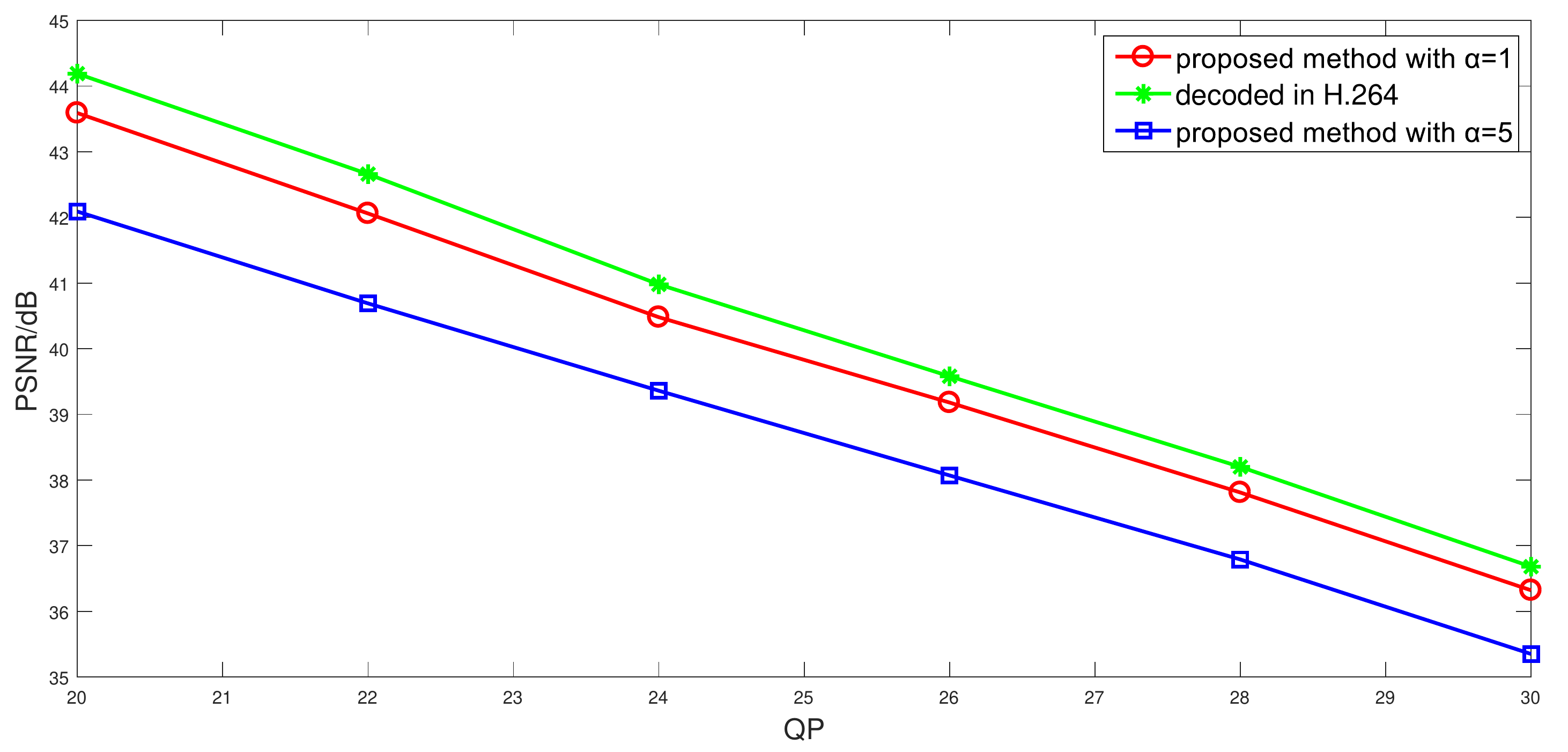}\hfill
\label{fig:subfigure1}}
\quad
\subfigure[akiyo]{%
 \includegraphics[width=.45\linewidth - 0.25mm]{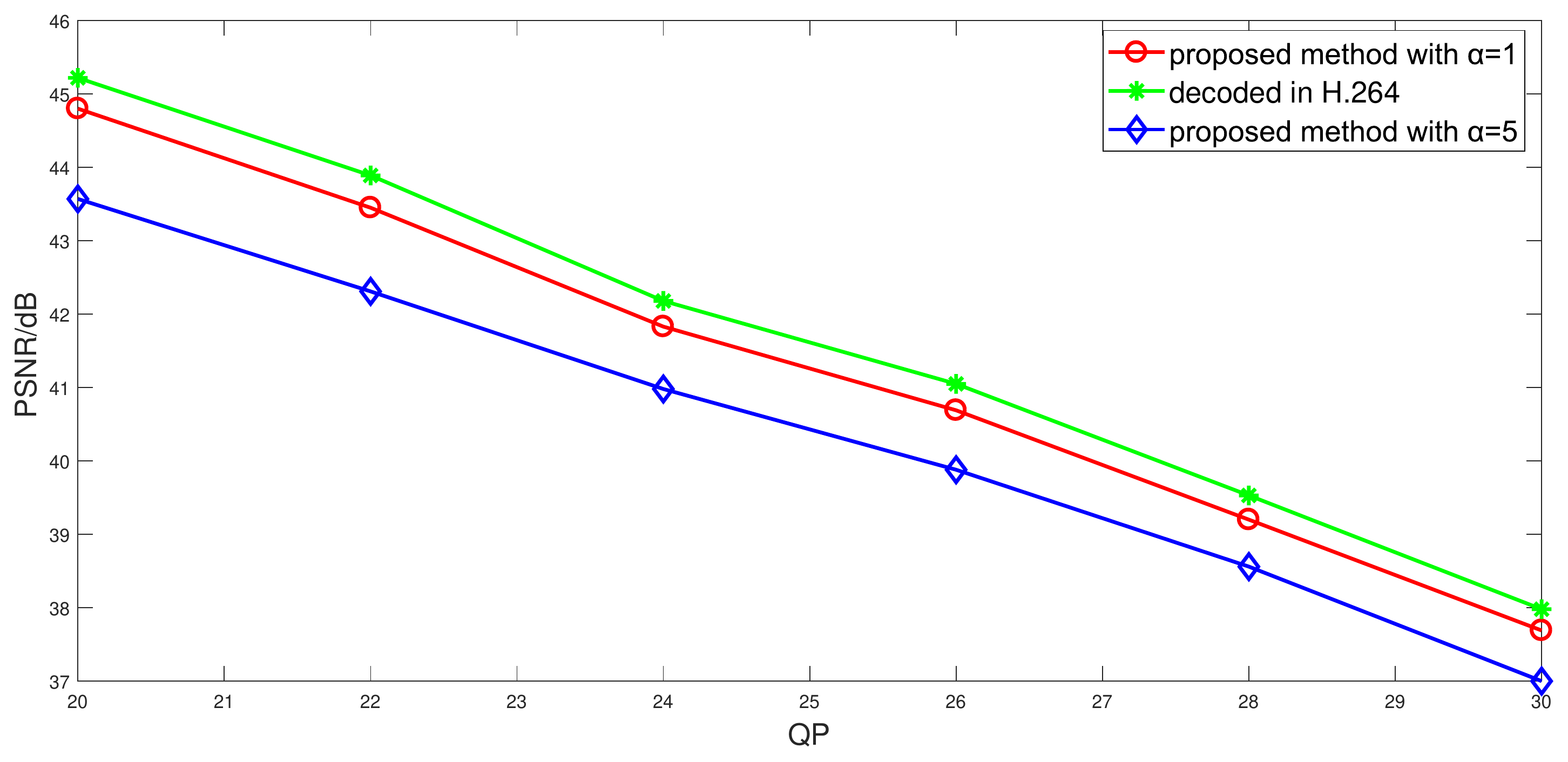}
\label{fig:subfigure2}}
\subfigure[hall]{%
\includegraphics[width=.45\linewidth - 0.25mm]{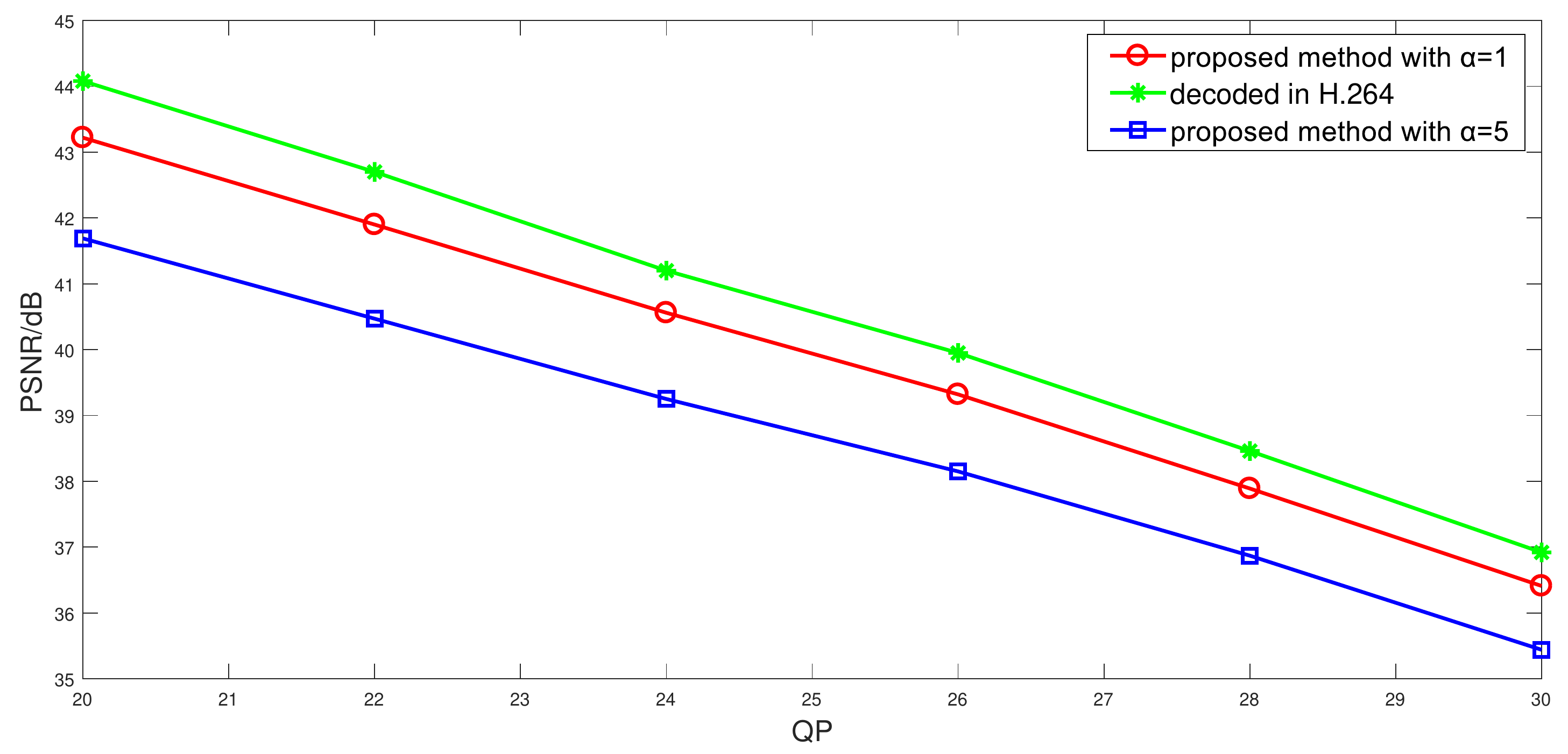}\hfill
\label{fig:subfigure3}}
\quad
\subfigure[foreman]{%
  \includegraphics[width=.45\linewidth - 0.25mm]{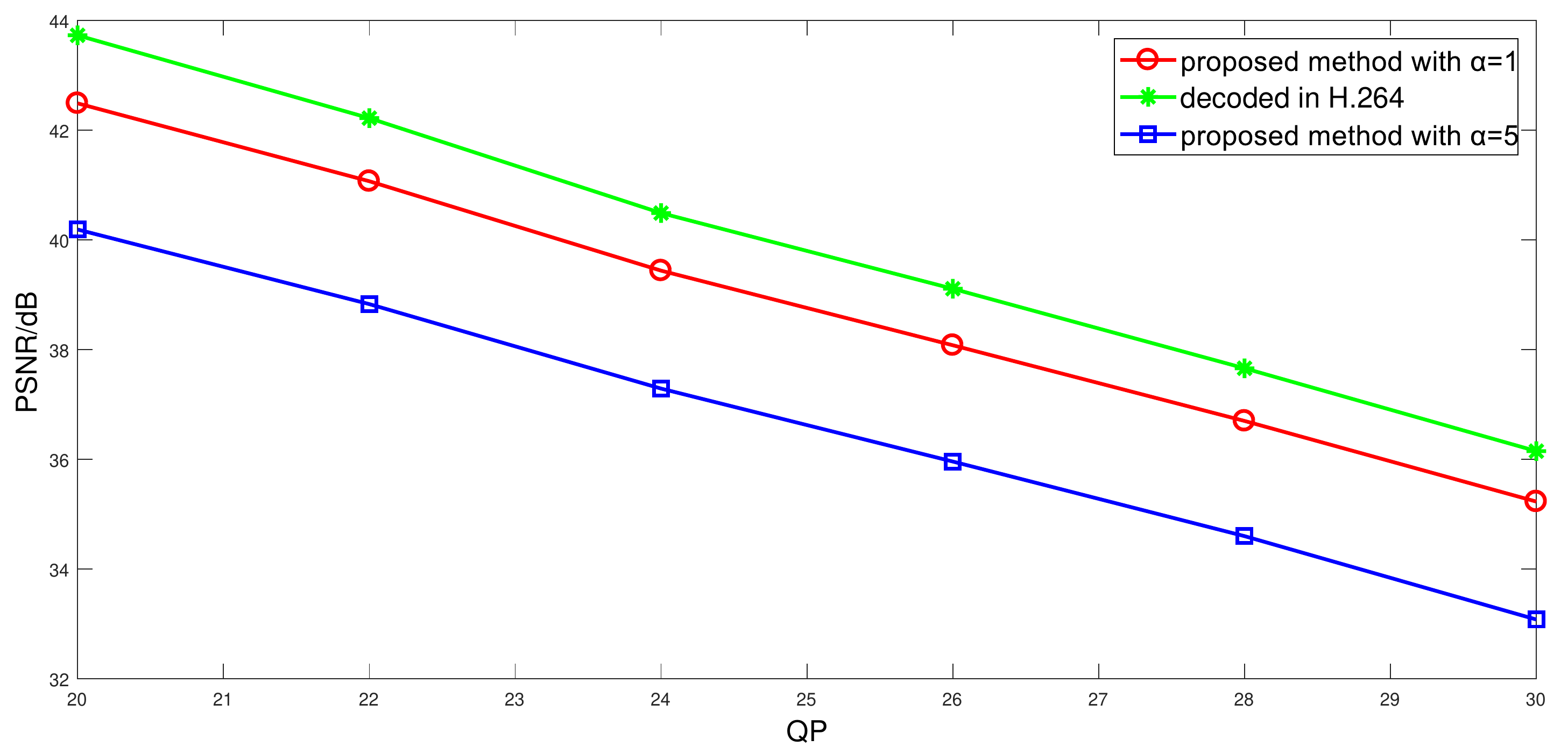}
\label{fig:subfigure4}}
\subfigure[silent]{%
\includegraphics[width=.45\linewidth - 0.25mm]{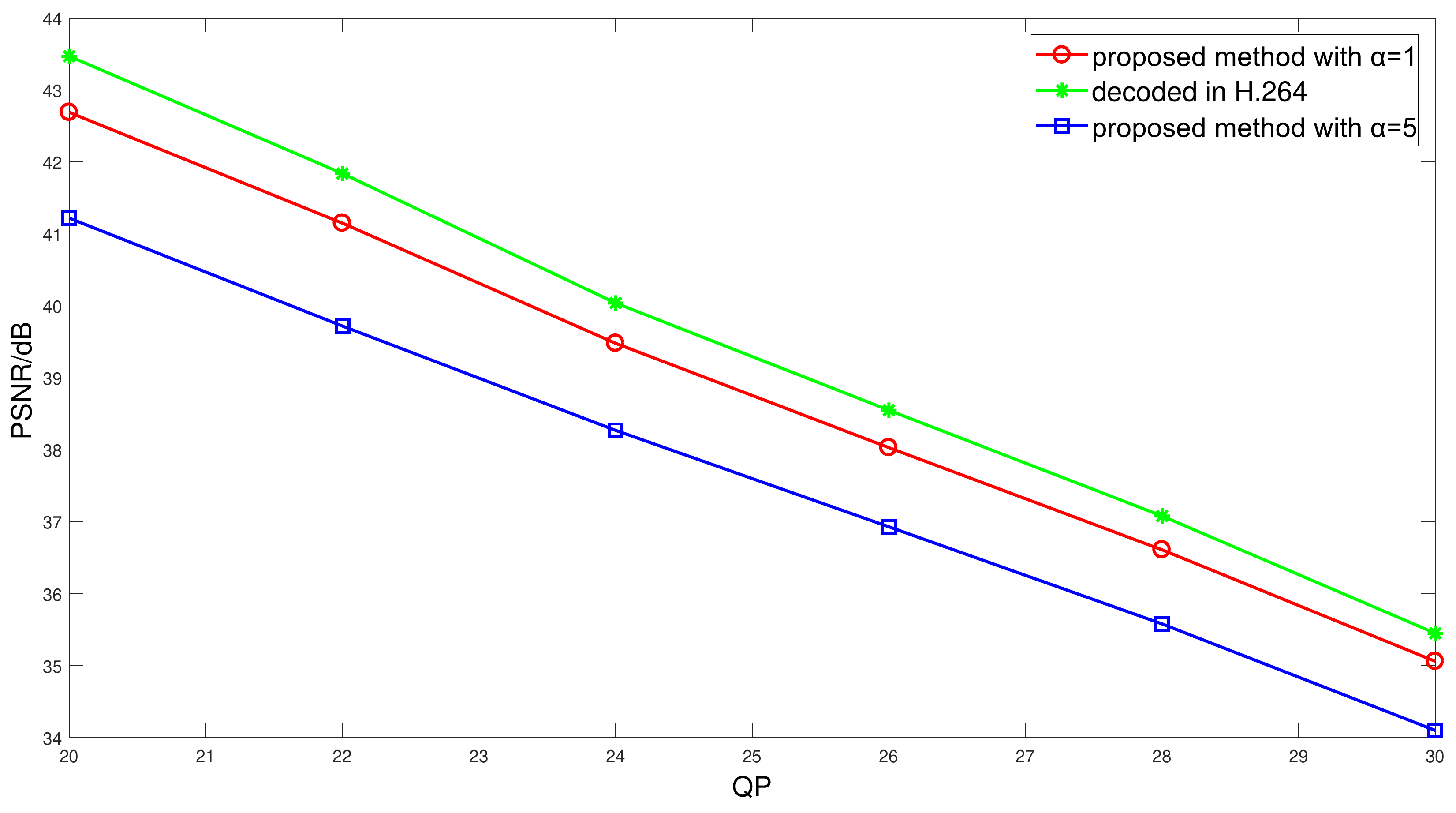}\hfill
\label{fig:subfigure1}}
\quad
\subfigure[coastguard]{%
 \includegraphics[width=.45\linewidth - 0.25mm]{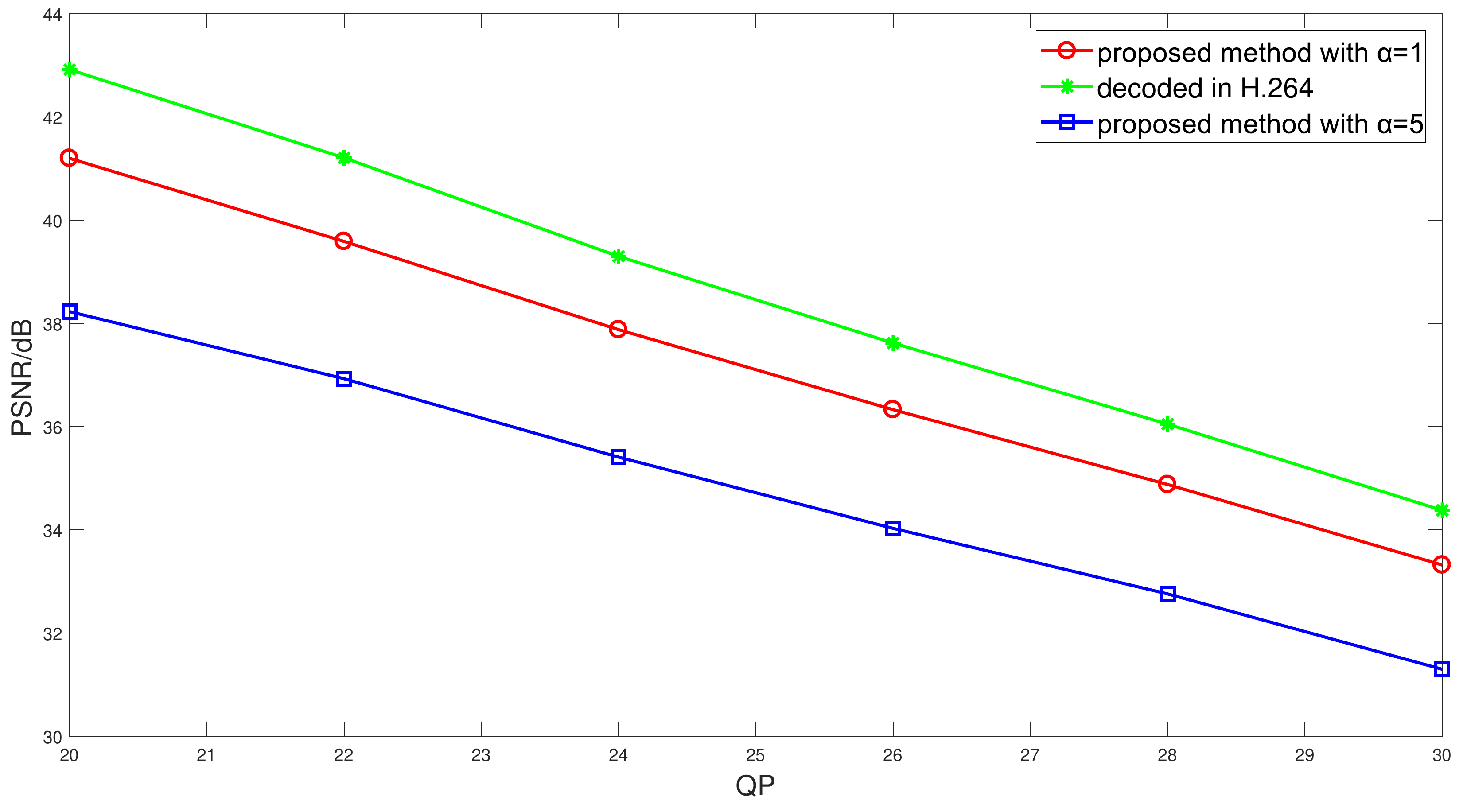}
\label{fig:subfigure2}}
\caption{he PSNR values comparison of marked frame with the limitation}
\label{markedpsnrcom}
\end{figure}

 Generally, at the decoder side, the quality of decoded frames without any other manipulations is the limitation of the concealed frames. If the peak signal to noise ratio (PSNR) values of marked frames are close to the limitations, the data hiding would have few influences on the marked video quality. However, the repeating frequency is also a key factor that influences the quality of the marked frames. Higher repeating frequency means more data is embedded into the frames. To indicate the concealment performance with different repeating frequencies, the cases with repeating frequencies $\alpha=1$ and $\alpha=5$ are compared with the limitation case. Fig.\ref{markedpsnrcom} shows the PSNR values comparison of marked frames with the decoded frames in the proposed method, and it shows the PSNR values with $\alpha=1$ is near the limitation.

\subsection{Performance of Error Concealment}

(1) Error concealment performance comparison

\begin{figure}[ht]
\centering
\subfigure[grandma]{%
\includegraphics[width=.45\linewidth - 0.25mm]{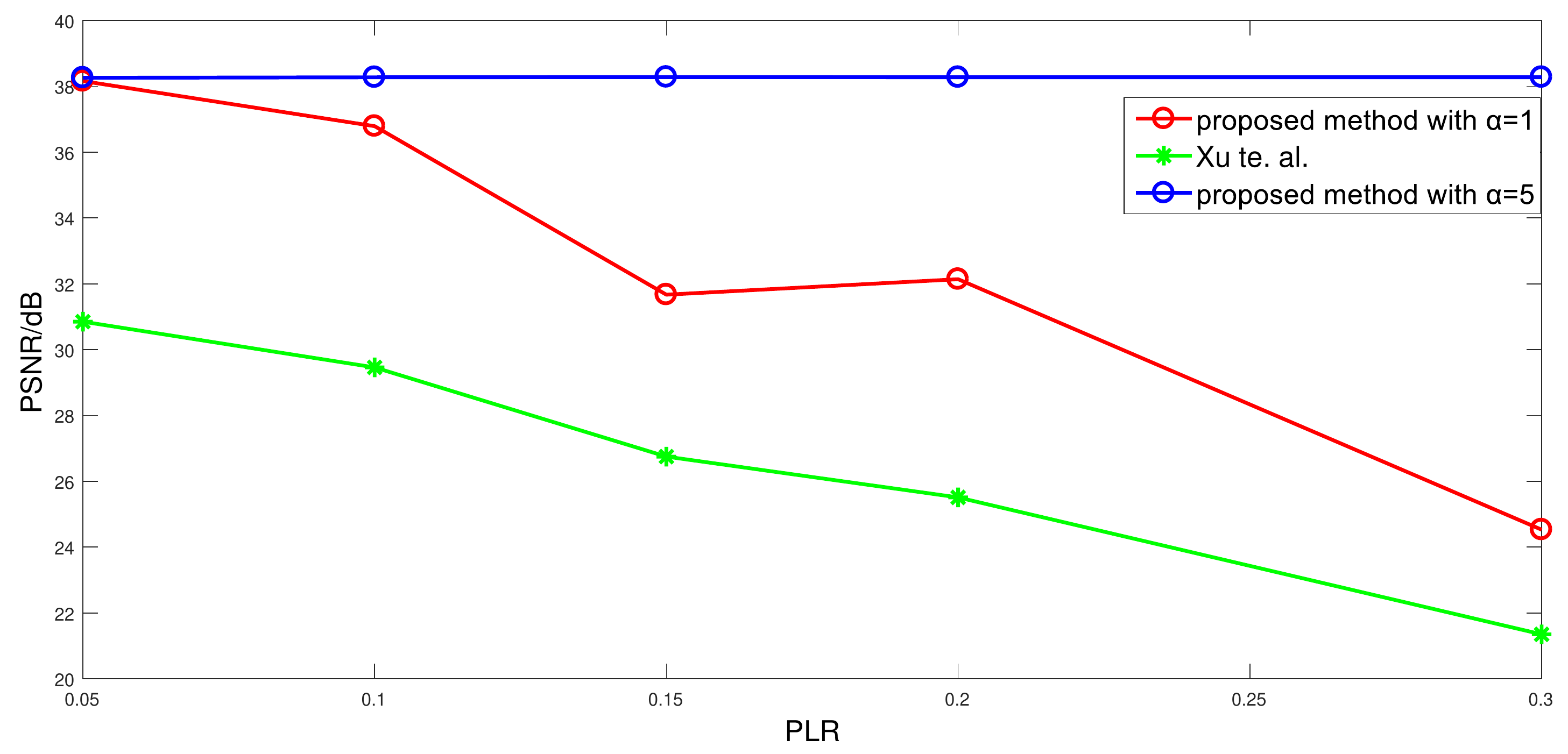}\hfill
\label{grandma}}
\quad
\subfigure[akiyo]{%
 \includegraphics[width=.45\linewidth - 0.25mm]{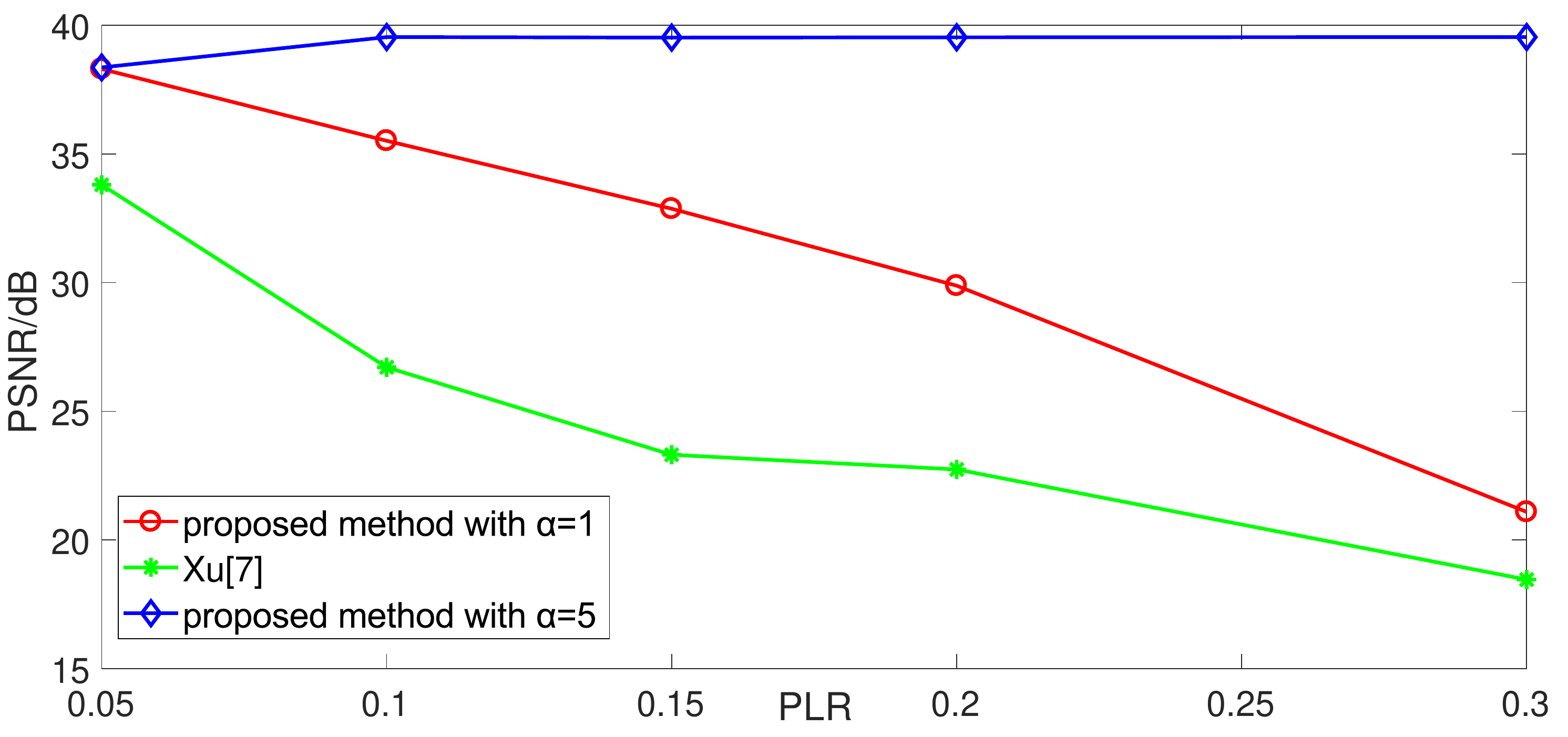}
\label{fig:akiyo}}
\subfigure[hall]{%
\includegraphics[width=.45\linewidth - 0.25mm]{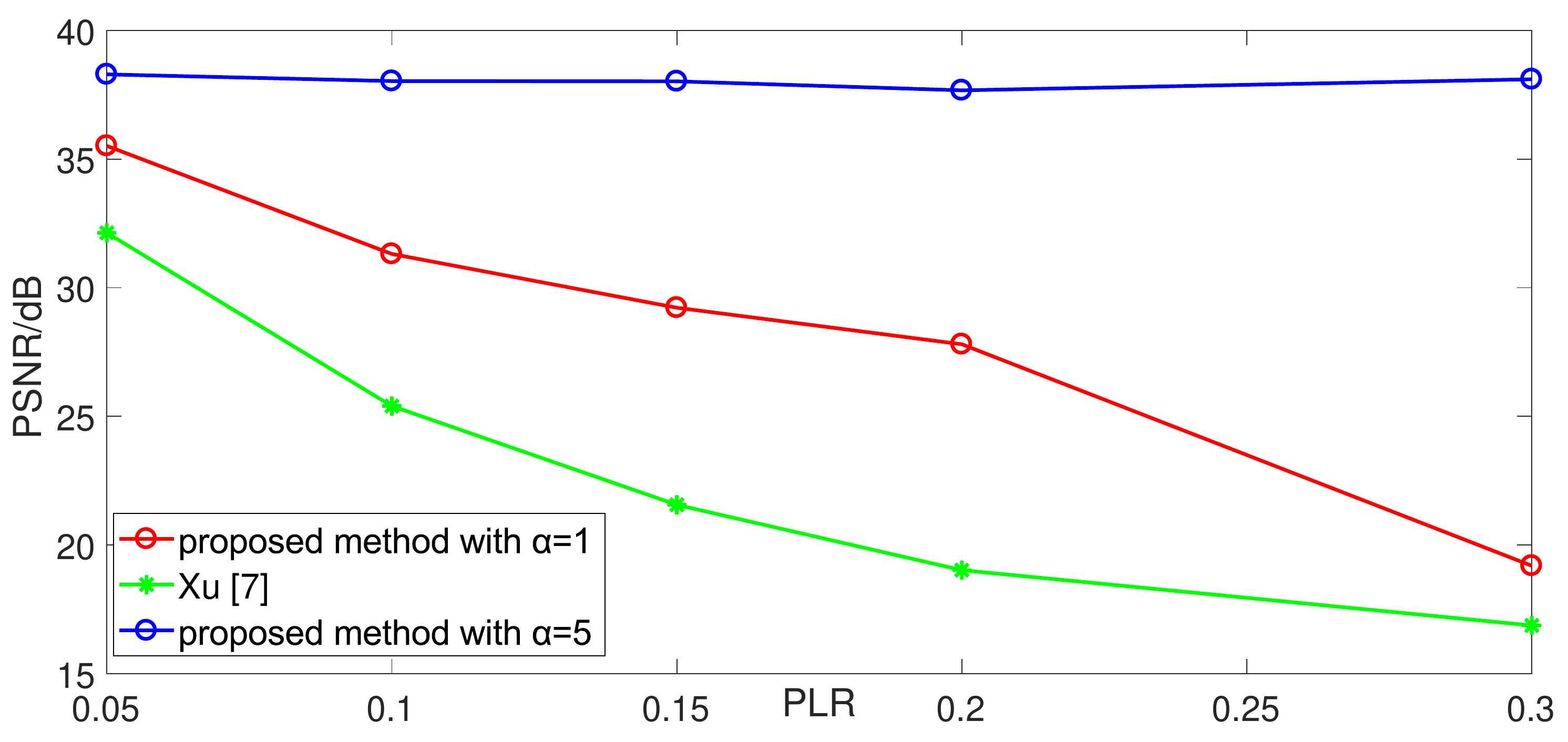}\hfill
\label{fig:hall}}
\quad
\subfigure[foreman]{%
  \includegraphics[width=.45\linewidth - 0.25mm]{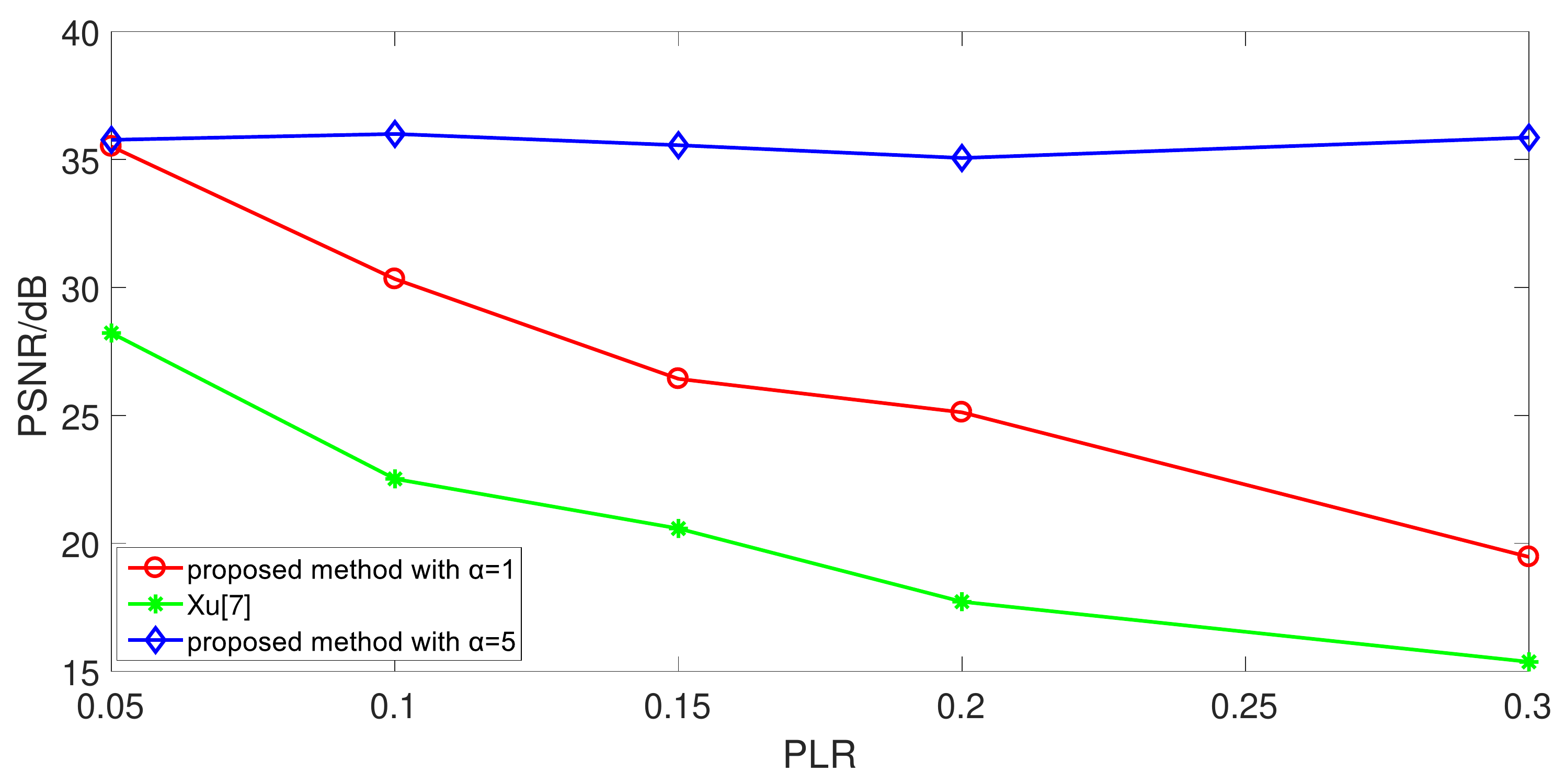}
\label{fig:foreman}}
\subfigure[silent]{%
\includegraphics[width=.45\linewidth - 0.25mm]{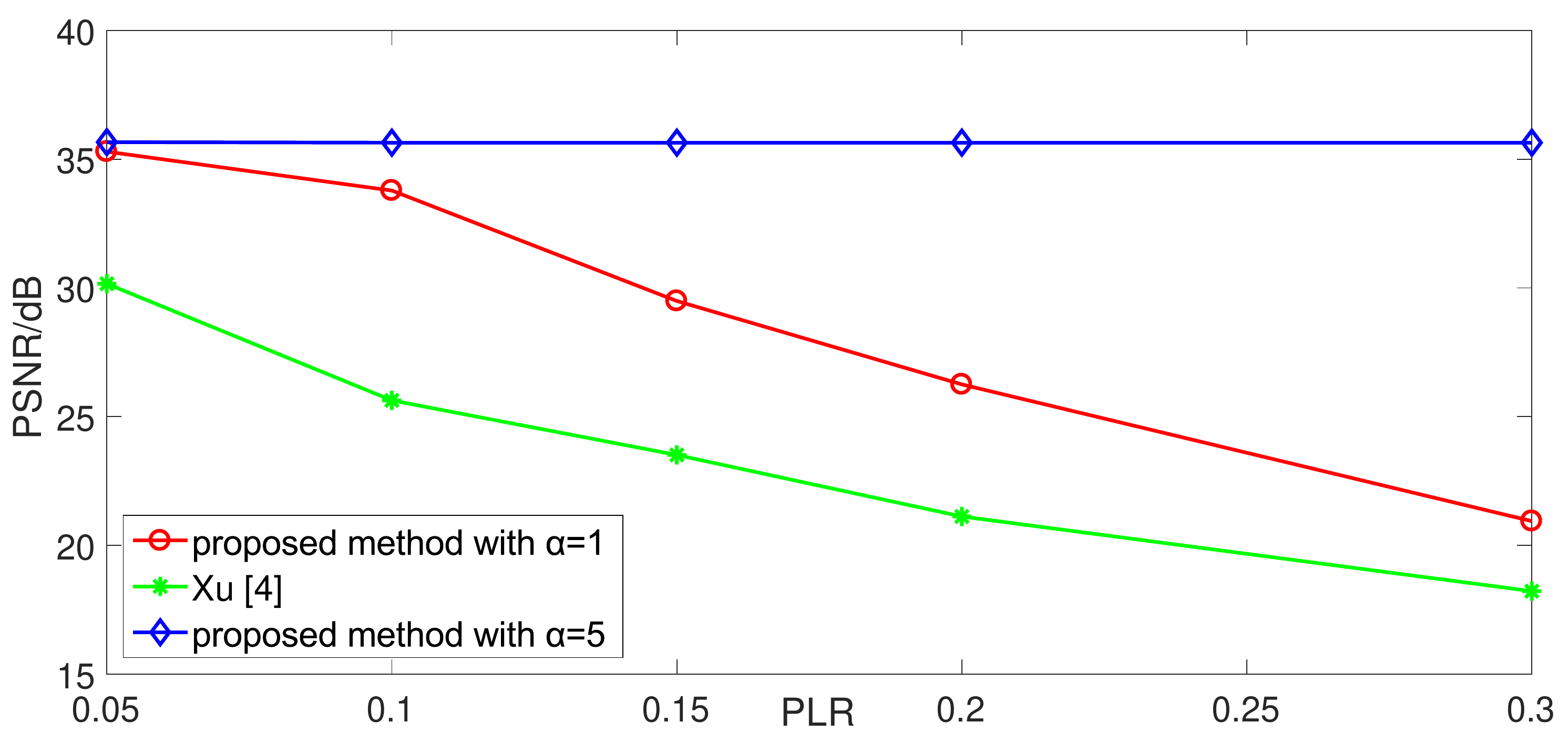}\hfill
\label{fig:silent}}
\quad
\subfigure[coastguard]{%
 \includegraphics[width=.45\linewidth - 0.25mm]{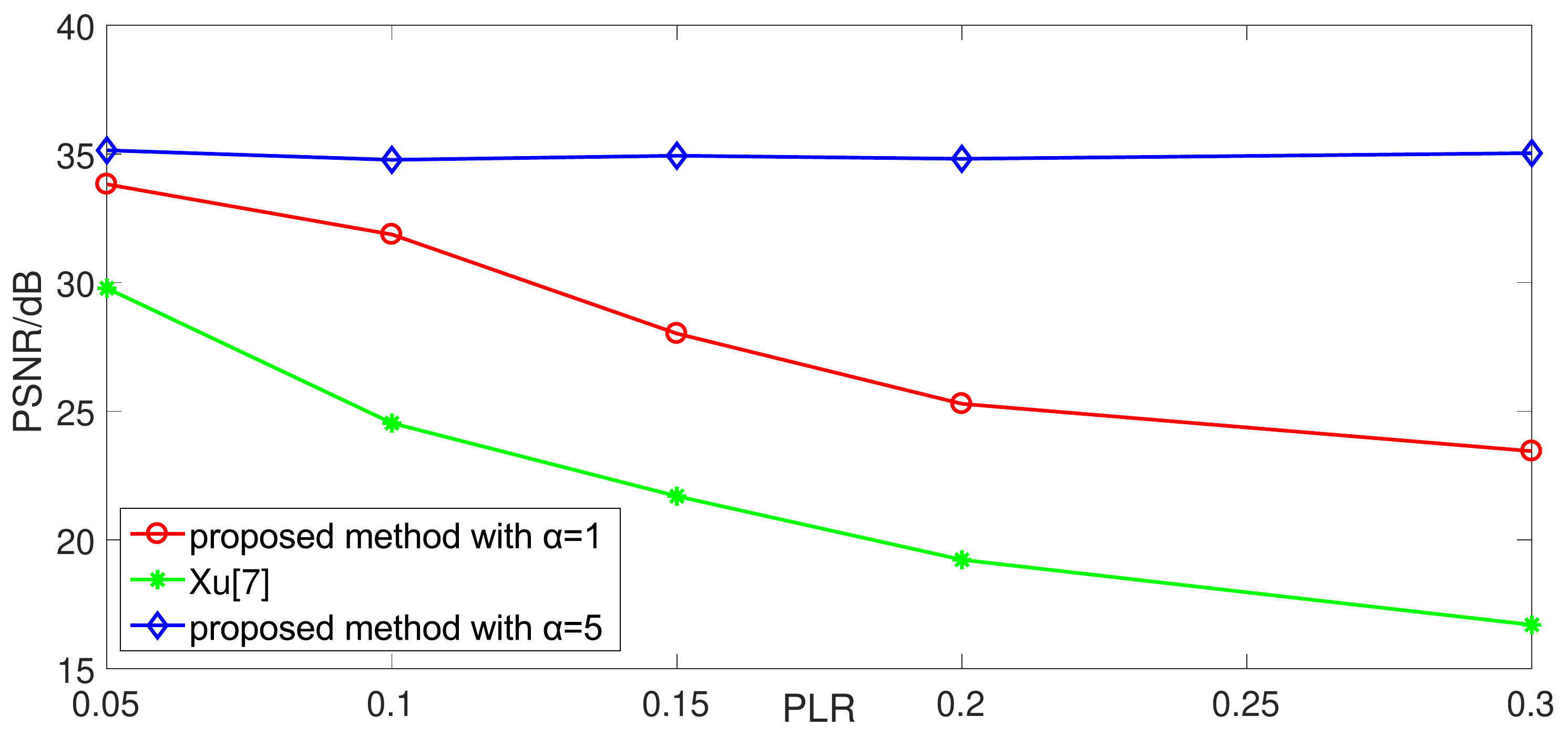}
\label{fig:coastguard}}
\caption{PSNR comparison of concealed frames}
\label{concealedpsnrcom}
\end{figure}

The error concealment performance is one of the most important features for this scheme.In cloud environment, data are stored in distributed way and may be from different servers. The packet lost rates (PLRs) can reflect all the channel states of the whole transition procedure. To simulate the transmission errors, different random intra-frame PLR are applied to each intra-frame, such as 0.05, 0.10, 0.15, 0.20 and 0.30. For every PLR, the performance is measured in terms of PSNR between original video sequence and concealed video sequence. Since \cite{R7} indicated the concealment performance in it is better than that in \cite{R15}, so the concealment performance comparison is taken with \cite{R7}. The PSNR comparisons of concealed frames are showed in Fig.\ref{concealedpsnrcom} and both the cases $\alpha=1$ and $\alpha=5$  are considered. According to Fig.8, the error concealment performance is better than that in \cite{R7} significantly. Especially, the corrupted frames are concealed completely with $\alpha=5$. The reason is that for the scheme in \cite{R7}, the cover MB for each MV is not random as the channel errors. In contrary, the embedding location of each MB is random in our proposed error concealment scheme, and also it reveal that Eq. (\ref{e6}) is correct.

\begin{figure}[ht]
\centering
\subfigure[original]{%
\includegraphics[width=.3\linewidth - 0.25mm]{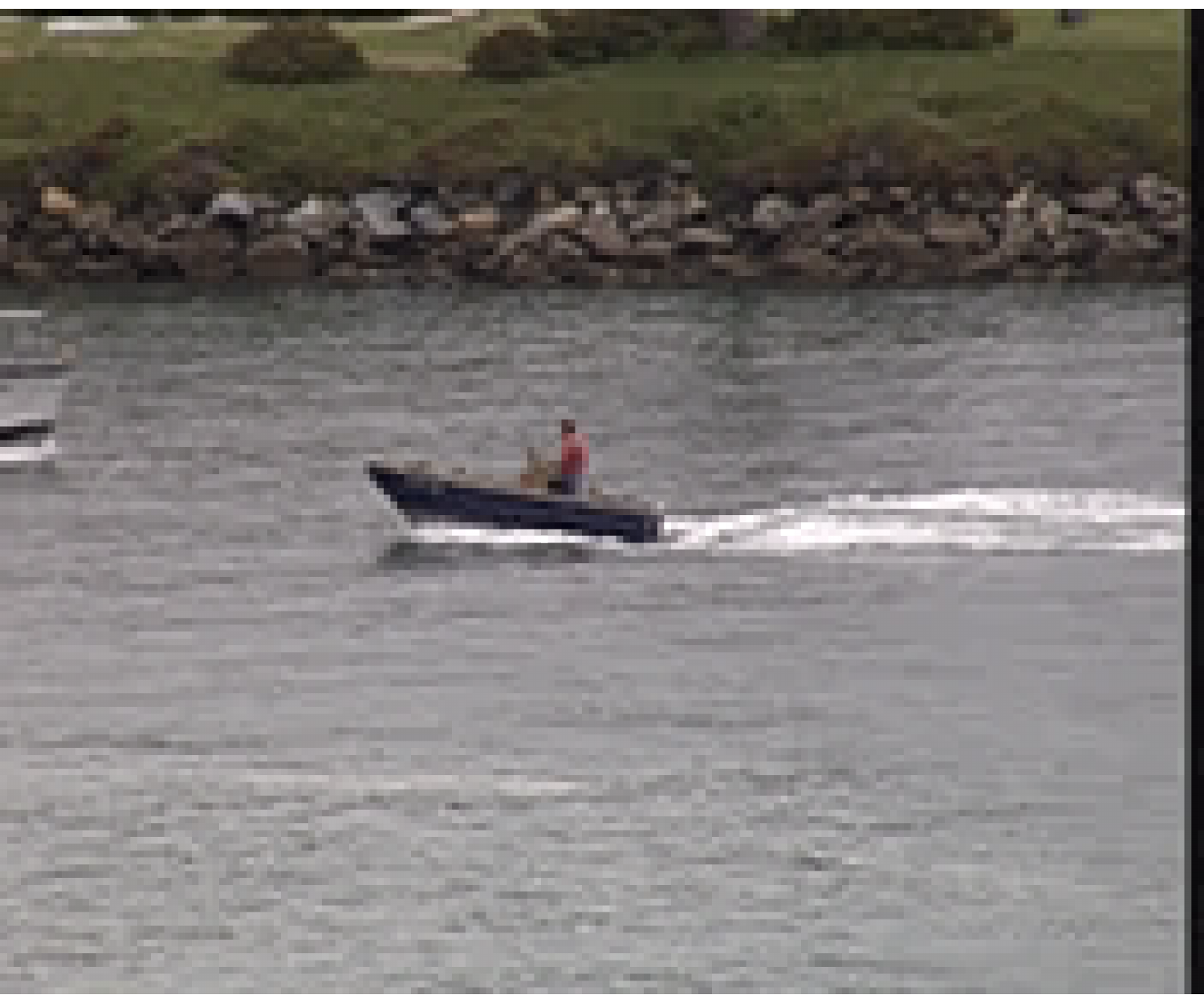}\hfill
\label{original frame}}
\subfigure[error]{%
 \includegraphics[width=.3\linewidth - 0.25mm]{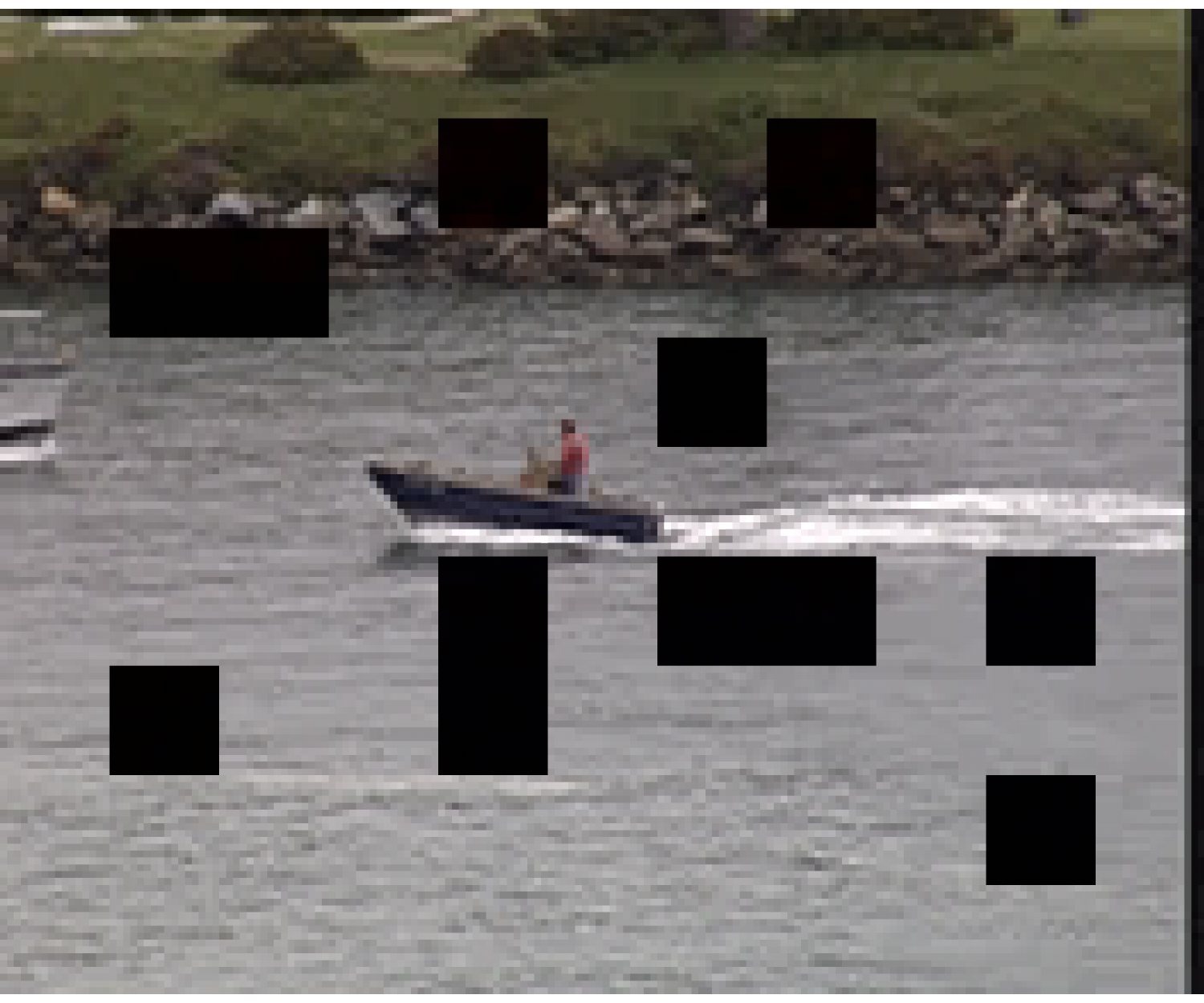}
\label{fig:cprrupted frame}}
\subfigure[conceaed]{%
\includegraphics[width=.3\linewidth - 0.25mm]{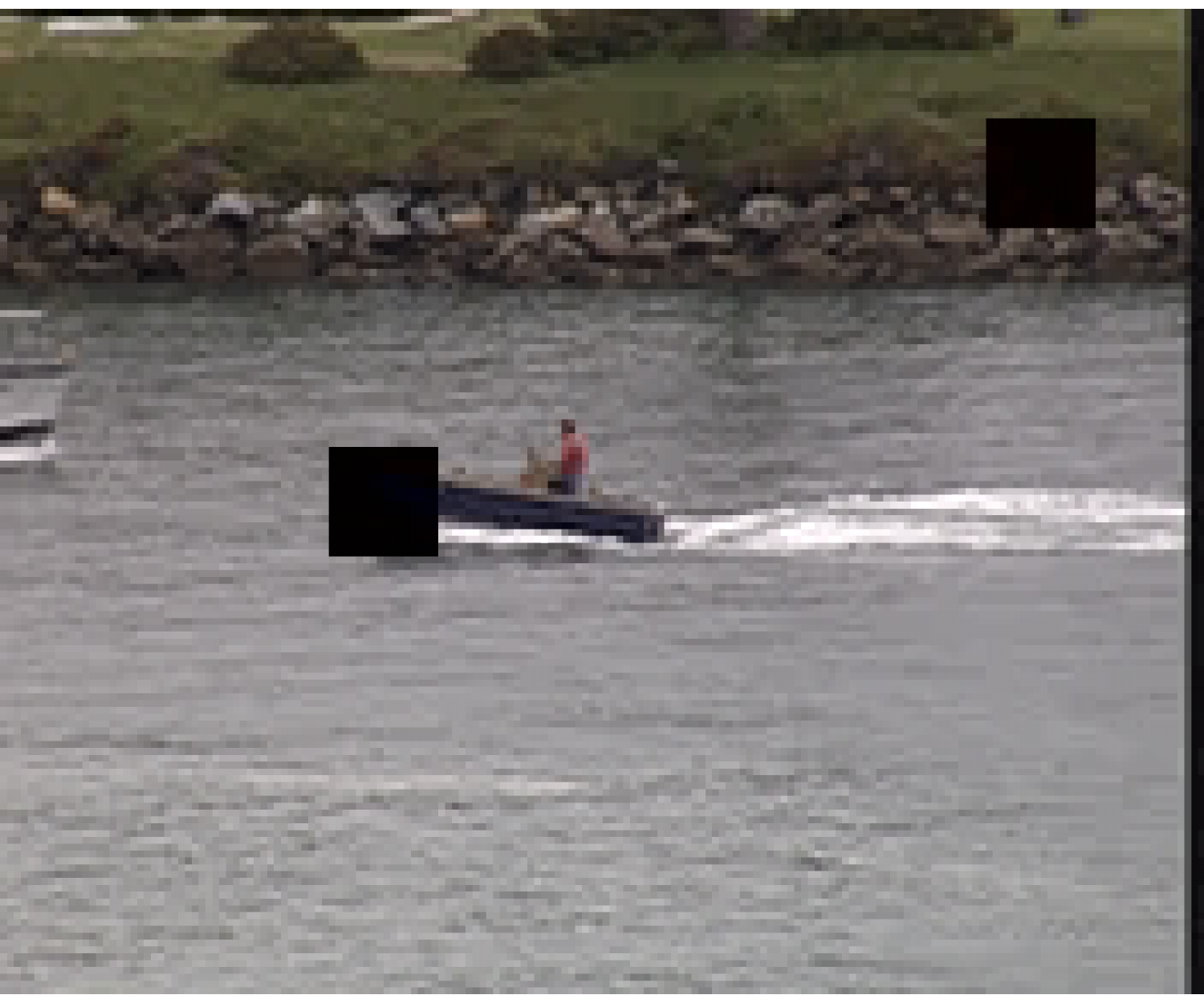}\hfill
\label{fig:concealed frame}}
\caption{Visual quality comparison for the \emph{coastguard} sequence with PLR 0.2 and $\alpha=1$}
\label{coastConcealed1}
\end{figure}

\begin{figure}[ht]
\centering
\subfigure[original]{%
\includegraphics[width=.3\linewidth - 0.25mm]{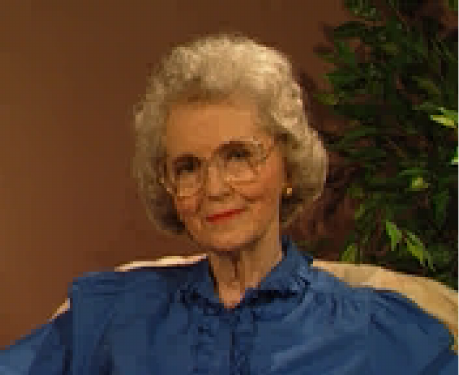}\hfill
\label{original frame}}
\subfigure[error]{%
 \includegraphics[width=.3\linewidth - 0.25mm]{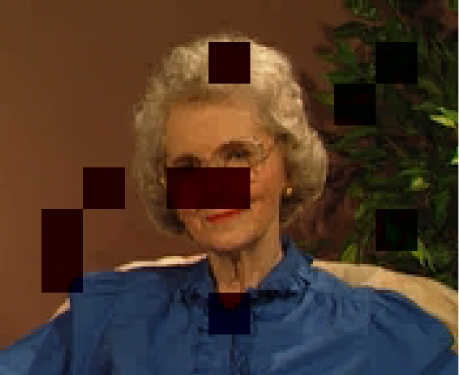}
\label{fig:cprrupted frame}}
\subfigure[conceaed]{%
\includegraphics[width=.3\linewidth - 0.25mm]{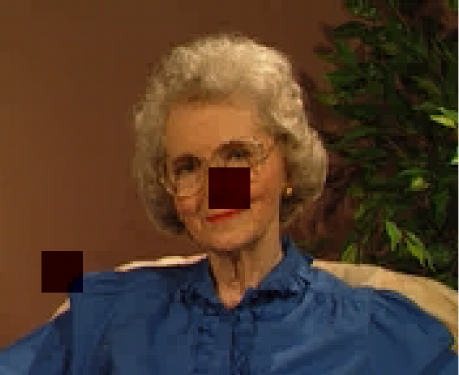}\hfill
\label{fig:concealed frame}}
\caption{Visual quality comparison for the \emph{grandma} sequence with PLR 0.2 and $\alpha=1$}
\label{grandConcealed1}
\end{figure}

\begin{figure}[ht]
\centering
\subfigure[original]{%
\includegraphics[width=.3\linewidth - 0.25mm]{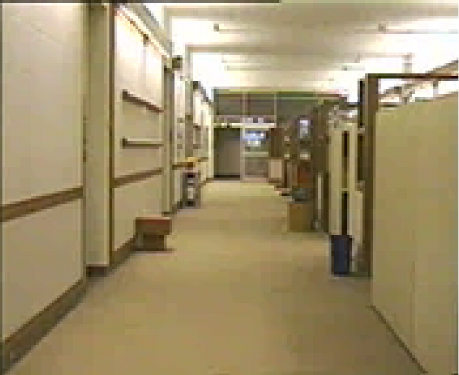}\hfill
\label{original frame}}
\subfigure[error]{%
 \includegraphics[width=.3\linewidth - 0.25mm]{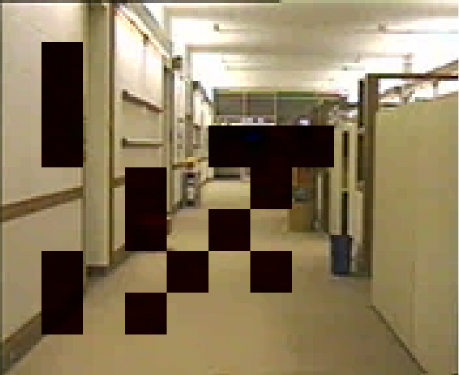}
\label{fig:cprrupted frame}}
\subfigure[concealed]{%
\includegraphics[width=.3\linewidth - 0.25mm]{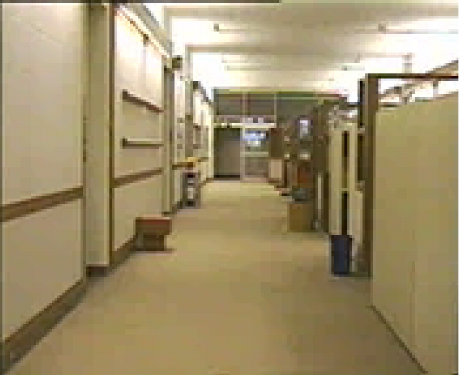}\hfill
\label{fig:concealed frame}}
\caption{Visual quality comparison for the \emph{coastguard} sequence with PLR 0.2 and $\alpha=1$}
\label{hallConcealed1}
\end{figure}

Fig.\ref{coastConcealed1}, Fig.\ref{grandConcealed1}, and Fig.\ref{hallConcealed1} shows the visual quality of original frame, error frame, concealed frame in ※coastguard§, ※grandma§ and ※hall§ sequences with PLR $0.2$ and $\alpha=1$ respectively. In this paper, all the errors caused by channel are occurred randomly and the corrupted frames are shown in Fig.\ref{coastConcealed1} (b), Fig.\ref{grandConcealed1} (b) and Fig.\ref{hallConcealed1} (b). The visual quality of the concealed frames is shown in Fig.\ref{coastConcealed1} (c), Fig.\ref{grandConcealed1} (c) and Fig.\ref{hallConcealed1}(c).

\begin{figure}[ht]
\centering
\subfigure[error]{%
 \includegraphics[width=.45\linewidth - 0.25mm]{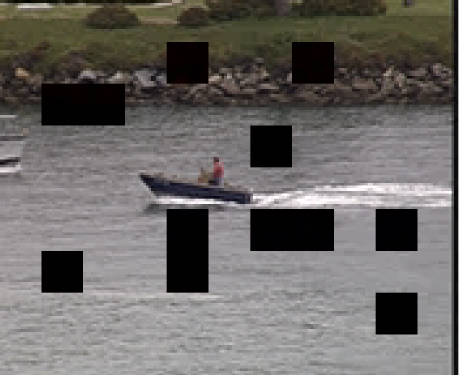}
\label{fig:cprrupted frame}}
\subfigure[conceaed]{%
\includegraphics[width=.45\linewidth - 0.25mm]{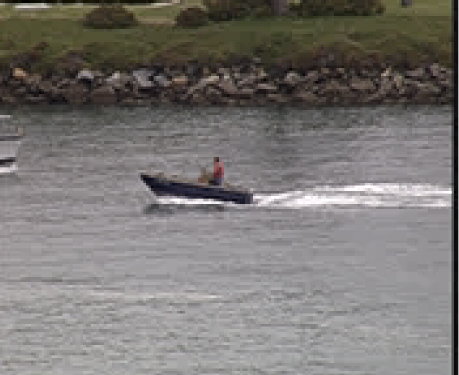}\hfill
\label{fig:concealed frame}}
\caption{Visual quality comparison for the \emph{coastguard} sequence with PLR 0.2 and $\alpha=5$}
\label{coastConcealed5}
\end{figure}

\begin{figure}[ht]
\centering
\subfigure[original]{%
\includegraphics[width=.45\linewidth - 0.25mm]{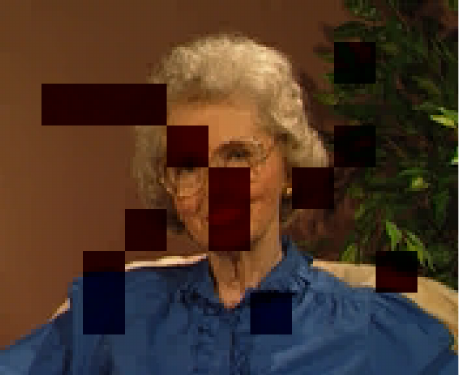}\hfill
\label{original frame}}
\subfigure[error]{%
 \includegraphics[width=.45\linewidth - 0.25mm]{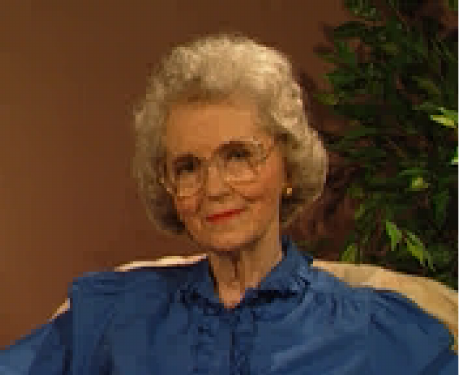}
\label{fig:cprrupted frame}}
\caption{Visual quality comparison for the \emph{grandma} sequence with PLR 0.2 and $\alpha=5$	}
\label{grandConcealed5}
\end{figure}

\begin{figure}[ht]
\centering
\subfigure[original]{%
\includegraphics[width=.45\linewidth - 0.25mm]{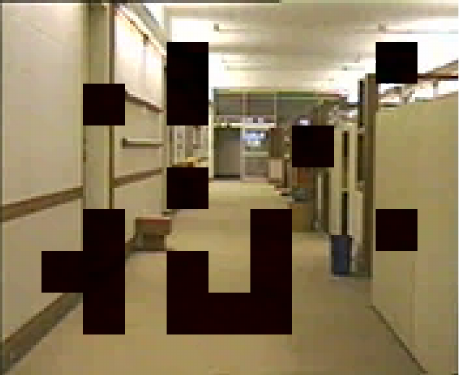}\hfill
\label{original frame}}
\subfigure[error]{%
 \includegraphics[width=.45\linewidth - 0.25mm]{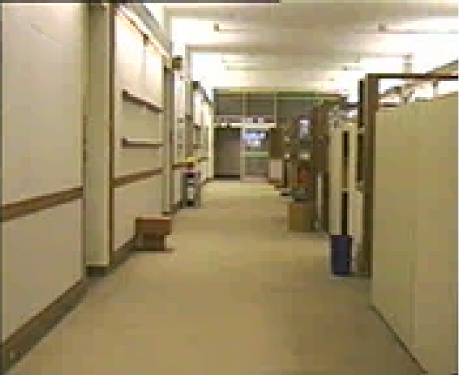}
\label{fig:cprrupted frame}}
\caption{Visual quality comparison for the \emph{hall} sequence with PLR 0.2 and $\alpha=5$}
\label{hallConcealed5}
\end{figure}

Furtherly, the visual quality of concealment performance is tested with $\alpha=5$. Fig.\ref{coastConcealed5} (a), Fig.\ref{grandConcealed5} (a) and Fig.\ref{hallConcealed5} (a) show the error frames, and their original frames are the same as the Fig.\ref{coastConcealed1}, Fig.\ref{grandConcealed1} and Fig.\ref{hallConcealed1}.  Fig.\ref{coastConcealed5} (b), Fig.\ref{grandConcealed5} (b) and Fig.\ref{hallConcealed5} (b) show the concealed frames with $\alpha=5$, the concealed frame is similar in visual to the original frames.

(2) MSE

PSNR denotes the peak SNR, it cannot reveal the distortion overall. Mean square error (MSE) is usually explored to capture the errors of every element, and it can reflect the comprehensive errors in a frame. For an error concealment scheme using RDH, the distortion may be generated both in the procedure of embedding and concealment. Without loss of generality, the original and the concealed frame with size $N_{1}\times N_{2}$ are denoted as $f_{o}$ and $f_{c}$, and the pixels in them are denoted as $P_{cij}, i=1,2,\cdots,N_{1},j=1,2,\cdots,N_{2}$, and $P_{oij}, i=1,2,\cdots,N_{1},j=1,2,\cdots,N_{2}$. The concealment distortion is defined as
 \begin{eqnarray}\label{e29}
 D_{cd}=\frac{1}{N_{1}\times N_{2}}\sum_{i,j} (P_{oij}-P_{cij})^2
 \end{eqnarray}

\begin{figure}[ht]
\centering
\subfigure[additivedisconceal]{%
\includegraphics[width=.45\linewidth]{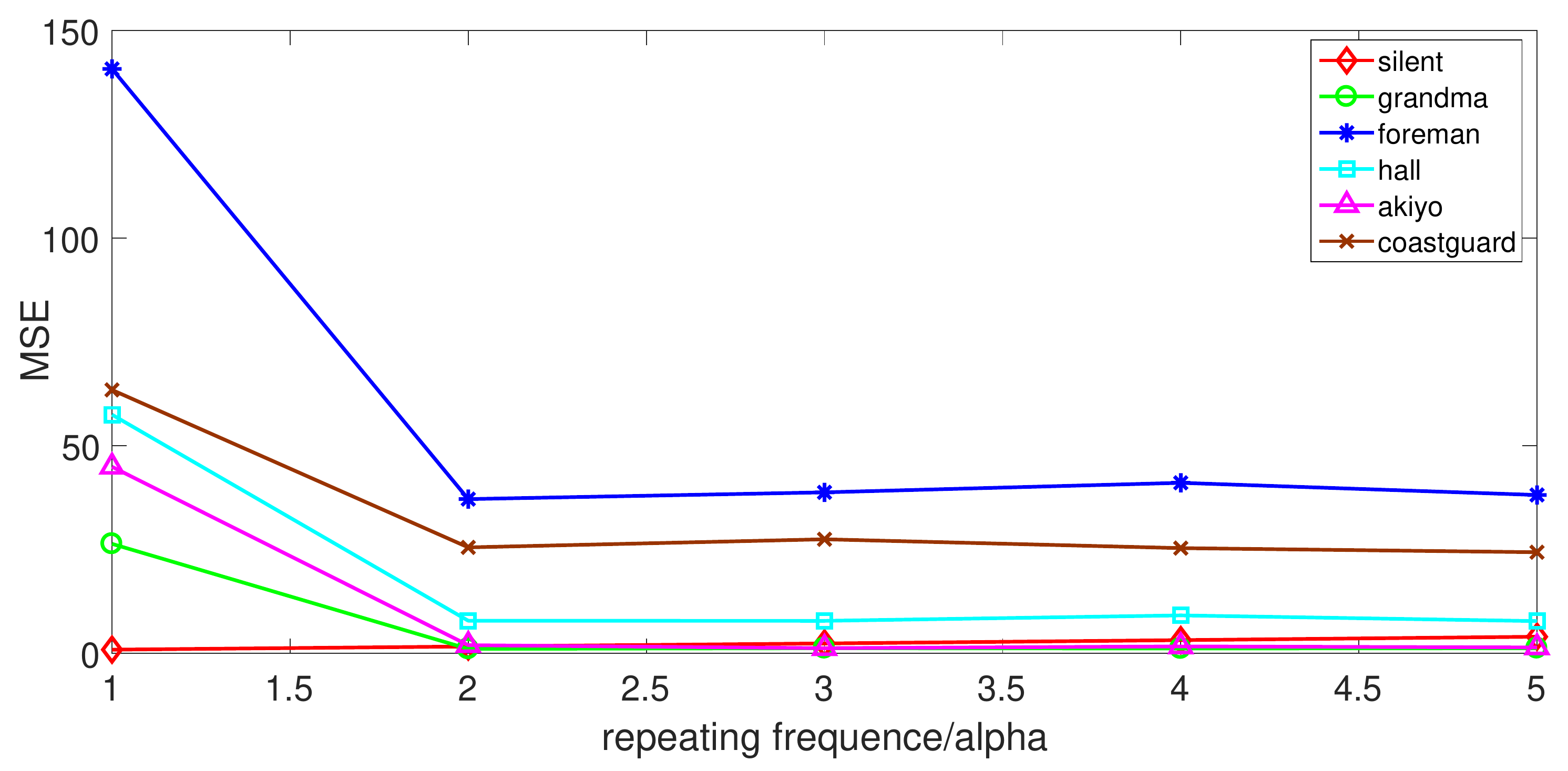}\hfill
\label{fig:PLR=0.2}}
\quad
\subfigure[foreman]{%
 \includegraphics[width=.45\linewidth ]{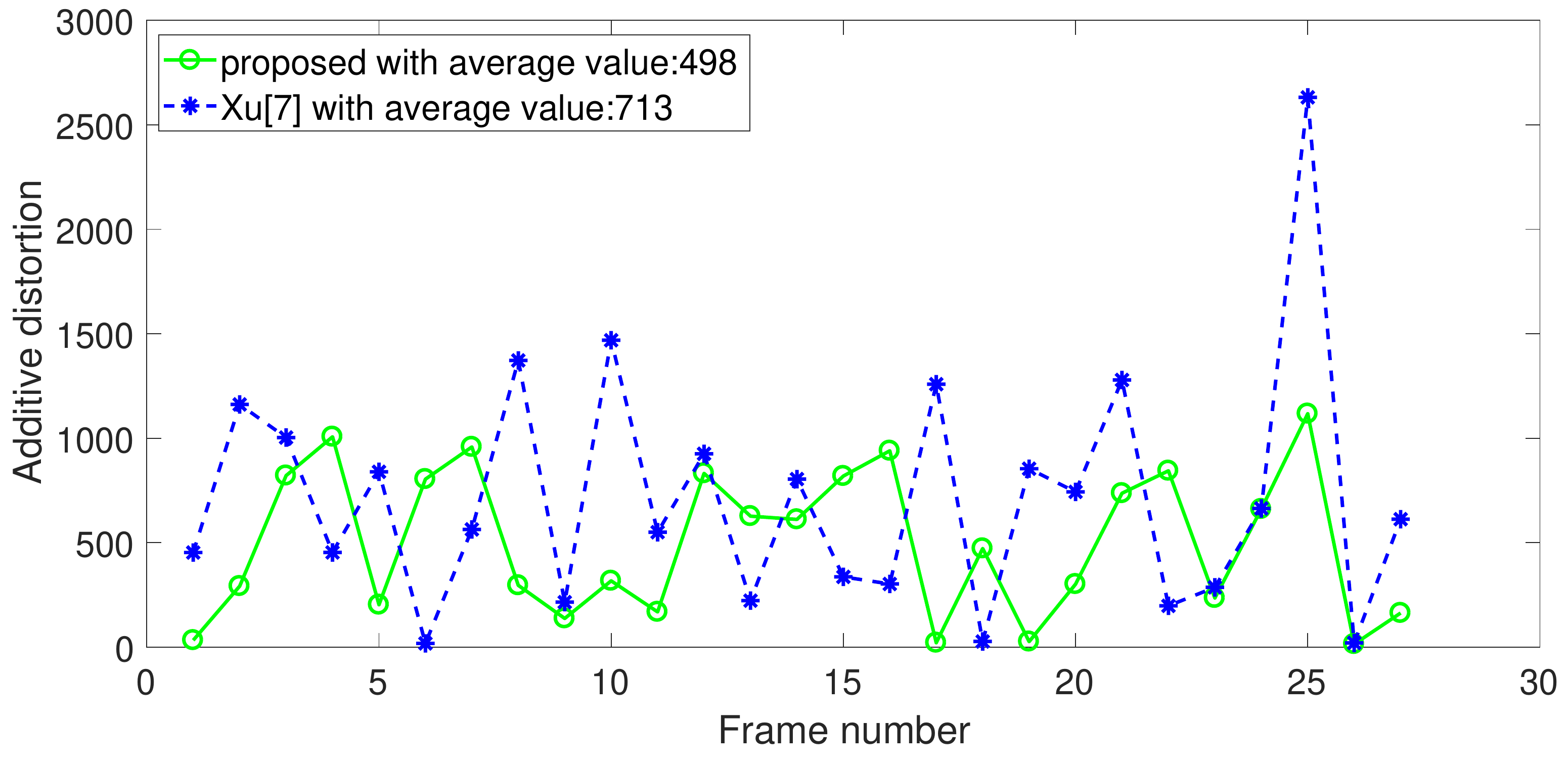}
\label{fig:foreman}}
\subfigure[hall]{%
\includegraphics[width=.45\linewidth ]{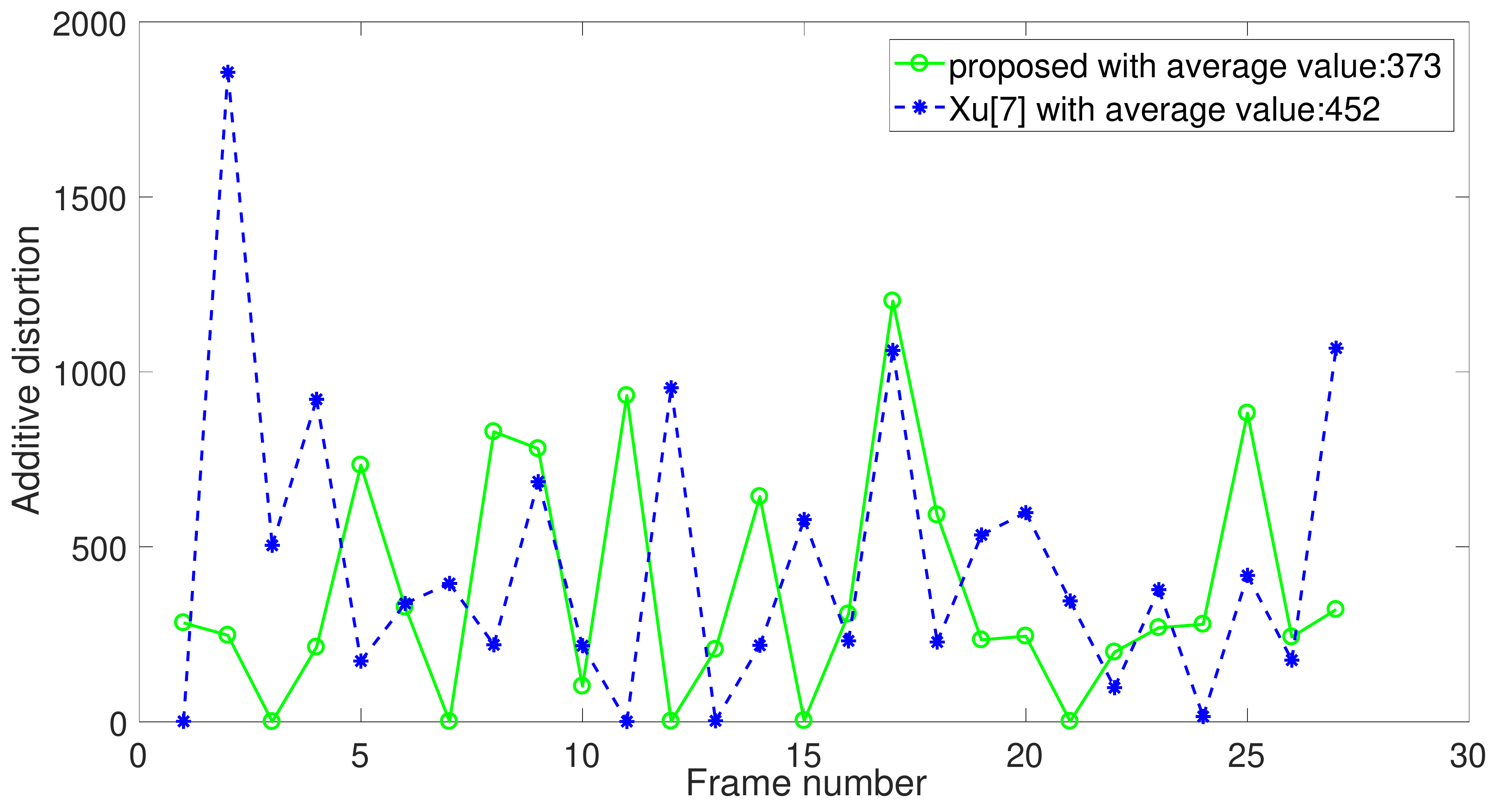}\hfill
\label{fig:hall}}
\quad
\subfigure[coastguard]{%
  \includegraphics[width=.45\linewidth ]{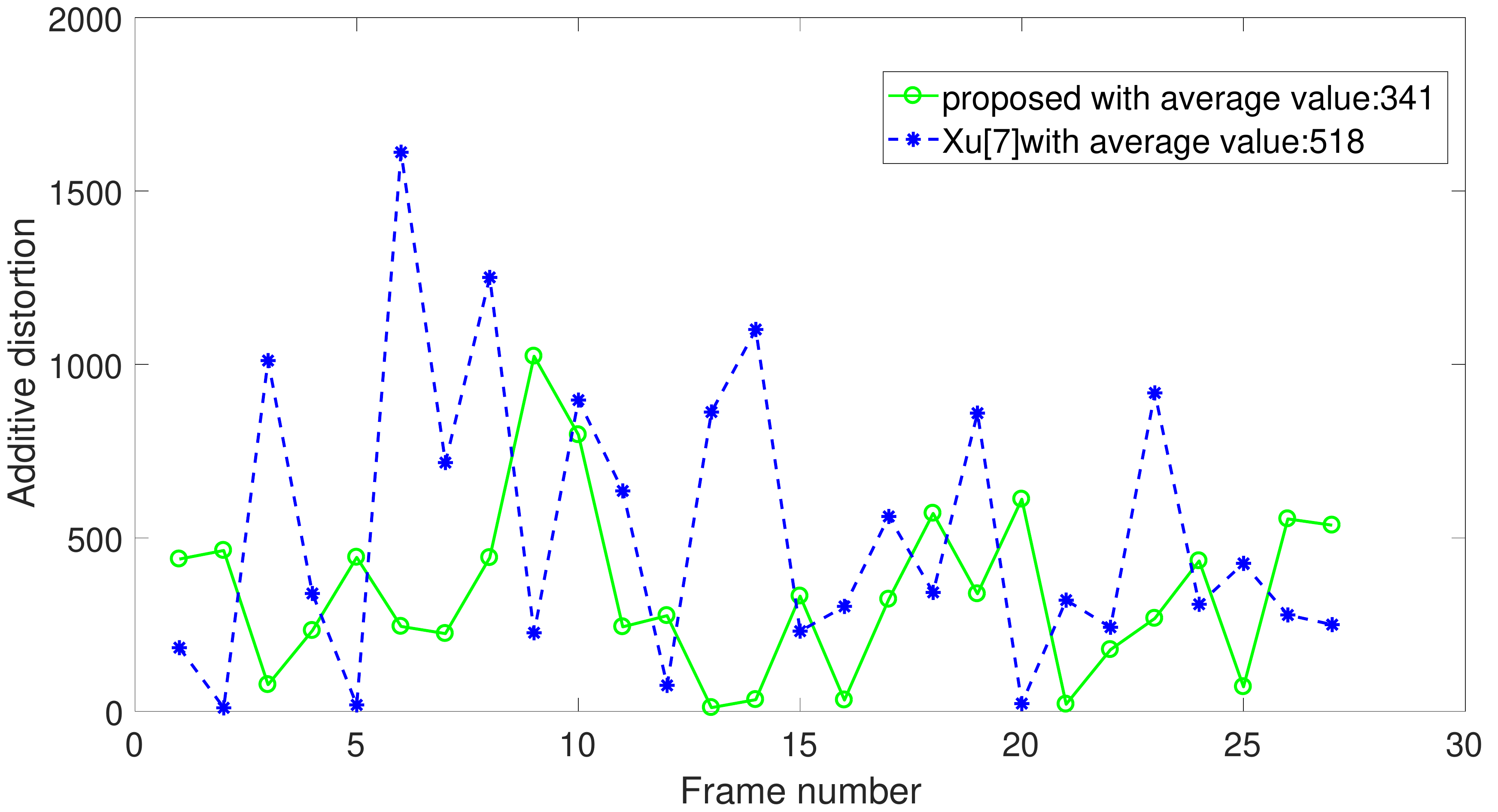}
\label{fig:coastguard}}
\caption{MSE of concealed frames with different PLR}
\label{mse}
\end{figure}

For the scheme in this paper, the quality of the concealed frames is controllable by adjusting the repeating frequency. However, the concealment performance is mainly decided by the error concealment method. As we can see, the PSNR values and the visual quality can illustrate the concealment performance, but they cannot reveal the detail distortions in a frame. The MSE is the summation of every distortion, also it can amplify the great distortion and reduce the low distortion. To explain the distortion of the concealment frames, Fig.\ref{mse} shows the MSE of concealed frames with different repeating frequencies and different videos. It shows MSE performance of almost all the video sequences will achieve a steady values with the repeating frequencies increasing. Also, for the test video ※foreman§, ＆hall＊, and ＆coastguard＊, the MSE in the proposed scheme is better than that in \cite{R7}. In addition, based on the extracted marked data, the concealment performance in this paper is mainly decided by the random embedding locations, the concealment performance may be different every time. And the Section 5.1 has explained the errors can be concealed with greater probability.
\section{Conclusion}
Reliable transmission of compressed video under mobile cloud environment is a challenging problem. In this paper, a 3D RDH-based video error concealment technique is proposed for H.264 codec. At the encode side, every three adjacent AC coefficients are counted to be a triple elements, and the data is embedded in the triples by shifting the triples. The proposed 3D RDH is utilized to embed the MVs into quantized DCT coefficients randomly. At decoder side, MVs are extracted and recovered to conceal the corrupted MBs. Meanwhile, the marked data is controllable by adjusting the MV repeating frequency, and both the concealment and embedding performance can be adjusted correspondingly. The experimental results show that the scheme is benefit from the repeated and random embedding locations, and also the proposed high capacity 3D RDH. By using the proposed 3D RDH method, rich space are utilized to embed more marked data. How to take advantage of the 3D space characteristic to exploit a more suitable histogram modifying should be researched in the future. Also, the error concealment scheme should be improved to the direction of few embedding bits and better concealment performance for the mobile cloud environment.

\section*{Acknowledgements}
%\begin{acknowledgements}
This work is supported by the National Natural Science Foundation of China (NSFC) under the grant No. U1536110.
%\end{acknowledgements}


%\end{IEEEbiography}


%\newpage



%\vfill


\end{document}